\documentclass[sigconf]{acmart}
\AtBeginDocument{%
  }
    
\usepackage{enumitem}
\usepackage{algorithm,bbding}
\usepackage{algorithmic,amsthm,amsfonts}
\usepackage{multirow}
\copyrightyear{2025}
\acmYear{2025}
\setcopyright{acmlicensed}\acmConference[KDD '25]{Proceedings of the 31st ACM SIGKDD Conference on Knowledge Discovery and Data Mining V.1}{August 3--7, 2025}{Toronto, ON, Canada}
\acmBooktitle{Proceedings of the 31st ACM SIGKDD Conference on Knowledge Discovery and Data Mining V.1 (KDD '25), August 3--7, 2025, Toronto, ON, Canada}
\acmDOI{10.1145/3690624.3709264}
\acmISBN{979-8-4007-1245-6/25/08}

\begin{document}

\title{Exploring Heterogeneity and Uncertainty for Graph-based Cognitive Diagnosis Models in Intelligent Education}

\author{Pengyang Shao}
\authornote{Equal Contribution.}
\orcid{0000-0003-2838-1987}
\affiliation{%
  \institution{Key Laboratory of Knowledge Engineering with Big Data, Hefei University of Technology}
  \city{Hefei}
  \state{Anhui}
  \country{China}
}

\author{Yonghui Yang}
\orcid{0000-0002-7601-6004}
\authornotemark[1]
\affiliation{%
  \institution{Key Laboratory of Knowledge Engineering with Big Data, Hefei University of Technology}
  \city{Hefei}
  \state{Anhui}
  \country{China}
}

\author{Chen Gao}
\orcid{0000-0002-7561-5646}
\affiliation{%
  \institution{BNRist, Tsinghua University}
  \city{Beijing}
  \country{China}
}

\author{Lei Chen}
\orcid{0000-0002-3193-7256}
\affiliation{%
  \institution{Department of Electronic Engineering, BNRist, Tsinghua University}
  \city{Beijing}
  \country{China}
}

\author{Kun Zhang}
\orcid{0000-0002-0743-9003}
\affiliation{%
  \institution{Key Laboratory of Knowledge Engineering with Big Data, Hefei University of Technology}
  \city{Hefei}
  \state{Anhui}
  \country{China}
}

\author{Chenyi Zhuang}
\orcid{0000-0003-1891-5666}
\affiliation{%
  \institution{Ant Group}
  \city{Hangzhou}
  \state{Zhejiang}
  \country{China}
}

\author{Le Wu}
\orcid{0000-0003-4556-0581}
\affiliation{%
  \institution{Key Laboratory of Knowledge Engineering with Big Data, Hefei University of Technology}
  \city{Hefei}
  \state{Anhui}
  \country{China}
}

\author{Yong Li}
\orcid{0000-0001-5617-1659}
\affiliation{%
  \institution{Department of Electronic Engineering, BNRist, Tsinghua University}
  \city{Beijing}
  \country{China}
}

\author{Meng Wang}
\orcid{0000-0002-3094-7735}
\authornote{Corresponding Author.}
\affiliation{%
  \institution{Key Laboratory of Knowledge Engineering with Big Data, Hefei University of Technology}
  \city{Hefei}
  \state{Anhui}
  \country{China}
}
\email{eric.mengwang@gmail.com}

\renewcommand{\shortauthors}{Pengyang Shao et al.}

\newtheorem*{myDef}{Problem Definition}
\newtheorem{myassumption}{Assumption}

\newcommand{\shortname}{ISG-CD}
\newcommand{\fullname}{\textbf{I}nformative \textbf{S}emantic-aware \textbf{G}raph-based \textbf{C}ognitive \textbf{D}iagnosis model}

\begin{abstract}
Graph-based Cognitive Diagnosis (CD) has attracted much research interest due to its strong ability on inferring students' proficiency levels on knowledge concepts.
While graph-based CD models have demonstrated remarkable performance, we contend that they still cannot achieve optimal performance due to the neglect of edge heterogeneity and uncertainty. Edges involve both correct and incorrect response logs, indicating heterogeneity. Meanwhile, a response log can have uncertain semantic meanings, e.g., \textit{a correct log can indicate true mastery or fortunate guessing, and a wrong log can indicate a lack of understanding or a careless mistake}. In this paper, we propose an \textbf{I}nformative \textbf{S}emantic-aware \textbf{G}raph-based \textbf{C}ognitive \textbf{D}iagnosis model (\textbf{\shortname}), which focuses on how to utilize the heterogeneous graph in CD and minimize effects of uncertain edges. 
Specifically, to explore heterogeneity, we propose a semantic-aware graph neural networks based CD model. To minimize effects of edge uncertainty, we propose an Informative Edge Differentiation layer from an information bottleneck perspective, which suggests keeping a minimal yet sufficient reliable graph for CD in an unsupervised way. We formulate this process as maximizing mutual information between the reliable graph and response logs, while minimizing mutual information between the reliable graph and the original graph. 
After that, we prove that mutual information maximization can be theoretically converted to the classic binary cross entropy loss function, while minimizing mutual information can be realized by the Hilbert-Schmidt Independence Criterion. 
Finally, we adopt an alternating training strategy for optimizing learnable parameters of both the semantic-aware graph neural networks based CD model and the edge differentiation layer. 
Extensive experiments on three real-world datasets have demonstrated the effectiveness of \shortname. 
\end{abstract}

\ccsdesc[500]{Applied computing~E-learning}

\keywords{Cognitive Diagnosis, Graph Neural Network, Student Modeling}


\maketitle

\section{Introduction} 

\begin{figure*}[t]
    \includegraphics[width=0.85\textwidth]{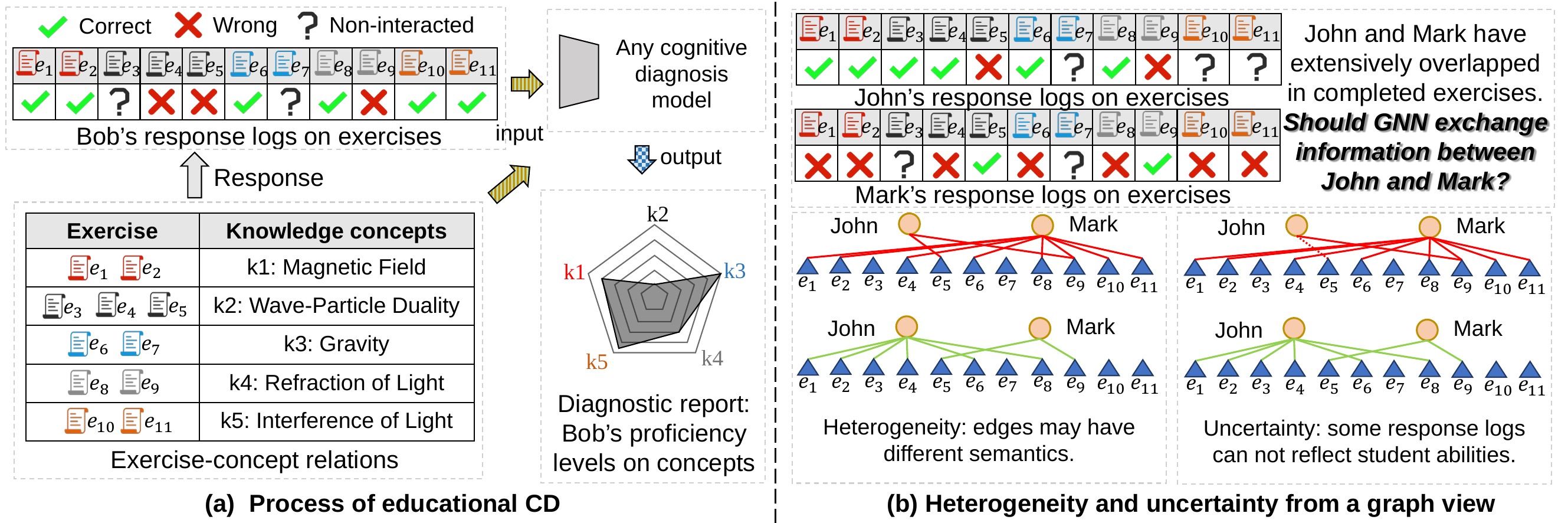}
    \caption{(a) The process of educational CD. These CD models take students' responses logs on exercises, and exercise-concept relations as input, and output students' proficiency levels on all knowledge concepts. 
     (b)  GNN between two students (John and Mark). Their interactions with exercises have extensively overlapped, however, they are not similar from a graph view due to heterogeneity. Further, there are uncertain edges, e.g., according to  interactions between John and exercises $e_3,e_4$, John has already grasped  concept $k2$. John's performance on exercise $e_5$ cannot reflect his ability on concept $k2$. }
    \label{fig:intro}
\end{figure*}

Educational CD has attracted much research interest due to growing needs in online education~\cite{embretson2013item,zhang2024path,wang2024survey}. 
Neural network-based CD models have garnered attention due to detailed diagnosis and relatively-high performance~\cite{wang2023dynamic,li2022hiercdf,ma2024enhancing,wang2024boosting,yang2024disengcd}. As shown in Figure \ref{fig:intro} (a), these models take students’ response logs, and expert-labeled exercise-concept relations as input, and output students’ proficiency levels on all concepts. The cornerstone lies in a diagnostic function that establishes connections between all trainable parameters and student-exercise response logs~\cite{wang2022neuralcd}.

Researchers have found that incorporating Graph Neural Network (GNN) into CD models can further improve diagnostic performance \cite{gao2021rcd,wang2023self}. 
Typically, response logs naturally form a student-exercise bipartite graph. Each node in the graph represents either a student or an exercise, and each edge represents a student-exercise response log. As GNNs have powerful feature learning and pattern extraction capabilities \cite{berg2017graph}, a natural idea is to adopt GNNs in capturing high-order graph information, thus enhancing the quality of student and exercise representations. 
These enhanced representations lead to more accurate predictions of students' proficiency levels on concepts.

While graph-based CD models have demonstrated remarkable performance, we contend that they still have inherent limitations \cite{gao2021rcd,wang2023self}. 
On one hand, edges may correspond to both correct and incorrect response logs, indicating edge \textbf{heterogeneity}. As shown in Figure \ref{fig:intro} (b), we consider: \textit{"Should GNN exchange information between John's and Mark's representations?"} 
As John and Mark have extensively overlapped in completed exercises, previous studies may exchange information between them~\cite{gao2021rcd,wang2023self}. However, they provide different answers to each exercise, indicating that their abilities are fundamentally opposite. Hence, we argue that GNN should not propagate information between them. 
On the other hand, each response log may have \textbf{uncertainty} in semantics, i.e., a correct log can indicate true mastery or fortunate guessing, and a wrong log can indicate a lack of understanding or a careless mistake. As shown in Figure \ref{fig:intro} (b), John has correctly answered {\small$e_3,e_4$} related to concept $k2$ but wrongly answered  {\small$e_5$} related to the same concept. It is hard to tell whether John's  response to {\small$e_5$} is due to a lack of ability or a careless mistake. 
We want to ensure that such uncertain response logs would not affect the process of GNN. 
Due to lack of labels, how to alleviate effects of uncertainty in an unsupervised way becomes our primary focus.

In this paper, we propose a novel \fullname~(\textbf{\shortname}), which focuses on how to utilize the heterogeneous graph in CD and minimize effects of uncertain edges.  Specifically, we take KaNCD as our backbone model due to its relatively high performance and easy-to-implementation~\cite{wang2022neuralcd}. KaNCD utilizes an embedding module that maps students, exercises, and concepts to a latent representation space. Then, the matrix factorization layer takes students' and concepts' latent representations as input, and outputs students' proficiency levels on concepts. Exercise difficulties can be obtained analogously. 
Compared to KaNCD, our improvements can be reflected in two aspects. First, to capture edge heterogeneity, we propose a \textbf{Semantic-aware GNN (S-GNN)} based CD model.  
Inspired by previous works \cite{berg2017graph}, we first obtain two subgraphs based on edge semantics (correct/incorrect response logs), and then update student and exercise embeddings based on these subgraphs. 
Second, to minimize effects of uncertain edges, we propose an \textbf{Informative Edge Differentiation (IE-Diff)} layer which can obtain a reliable student-exercise graph. 
Due to the lack of uncertainty labels, IE-Diff layer is designed based on  Information Bottleneck (IB) principle~\cite{saxe2019information,tishby2015deep}, which can keep a minimal yet sufficient reliable graph for CD in an unsupervised way.   
We parameterize the reliable graph based on the probability of dropping edges. Obviously, the reliable graph should assign high probabilities to those uncertain edges, and assign low probabilities to certain edges. 
To model the probability of dropping edges, we use Bernoulli distributions, whose parameters are estimated from corresponding student and exercise latent embeddings. 
Then, we summarize the goal of IE-Diff as maximizing mutual information between the reliable graph and response logs, while minimizing mutual information between the reliable graph and the original graph. 
Directly minimizing/maximizing mutual information is difficult, therefore, we prove that mutual information maximization between the reliable graph and response logs can be theoretically converted to the classic binary cross entropy loss function, while minimizing mutual information between the reliable graph and the original graph can be realized by the Hilbert-Schmidt Independence Criterion~\cite{gretton2005measuring,ma2020hsic}. 
Finally, we adopt an alternating training strategy to together optimize learnable parameters of the S-GNN based CD model and the IE-Diff layer~\cite{zibetti2022alternating,zhu2024collaborative}. Extensive experiments on three real-world datasets have demonstrated the effectiveness of \shortname. 
The major contributions can be summarized as follows: 
\begin{itemize}[leftmargin=0.3cm]
\setlength{\itemsep}{0pt}
\setlength{\parskip}{0pt}
\item We explore semantic heterogeneity and uncertainty in the student-exercise bipartite graph. 
We argue that neglecting heterogeneity and uncertainty will lead to sub-optimal diagnostic performance of graph-based CD models. 
\item We propose a novel  \shortname~model, which focuses on how to utilize the heterogeneous graph and minimize effects of uncertain edges. We first propose a semantic-aware GNN layer to utilize the graph, and then propose an informative edge differentiation layer to handle uncertainty. 
\item Extensive experiments on three real-world datasets have demonstrated the stable effectiveness of \shortname, e.g., \shortname~ achieves stable improvements of over 1\% than best baselines on the DOA metric on all three datasets. 
 \end{itemize}

\section{Related Work}
\subsection{Educational Cognitive Diagnosis}
Research on educational CD has primarily emerged from the field of psychology. Two classic methods in this area are Item Response Theory (IRT)~\cite{embretson2013item} and DINA~\cite{de2009dina}. 
IRT utilizes continuous one dimensional parameters to represent student and exercise entities. Further, it employs a logistic function to infer students' performance on exercises. Multidimensional Item Response Theory (MIRT) has been proposed for better ability estimations by extending parameter dimensions~\cite{reckase1997past}. 
DINA represents student and exercise entities with binary discrete variables. Then, it introduces ``slip'' and ``guess'' parameters to better fit real-world scenarios~\cite{de2009dina}. 
With the recent advancements in neural networks, researchers have turned their attention to leveraging neural networks in educational CD~\cite{gao2023leveraging,cui2024leveraging,liu2024inductive,yang2024disengcd,zhang2024understanding}. NCDM framework first utilizes high-dimensional continuous representations to model students' abilities and exercise difficulties and proposes a novel diagnostic function to incorporate all parameters to obtain final results~\cite{wang2020neural}. 
Inspired by collaborative filtering~\cite{cai2024popularity,chen2023improving,zhang2023fairlisa}, KaNCD utilizes a matrix factorization layer to fuse students' and concepts' latent representations for students' proficiency levels on concepts~\cite{wang2022neuralcd}. 

Building upon the success of the NCDM framework, graph-based CD models integrate graph structure information into CD models~\cite{ma2024dgcd,qian2024orcdf,shen2024capturing}. RCD simultaneously captures graph information hidden in the student-exercise bipartite graph and concept relationships~\cite{gao2021rcd}. 
Compared to RCD, SCD introduces a graph self-supervised learning to boost student prediction performance~\cite{wang2023self}. 
ORCDF further proposes a response-aware GNN layer with a graph contrastive learning based regularization term to enhance the robustness of graph-based CD models~\cite{qian2024orcdf}. 
Compared to these graph-based CD models, our distinction lies in exploring the semantic edge heterogeneity and minimizing effects of uncertain edges.

\subsection{Uncertainty Detection in GNN}
Researchers point out that there may exist uncertainty in a graph structure, leading to suboptimal performance of GNN. An intuitive approach is to adaptively select node features or edges ~\cite{liu2022agfa,chen2023adap,zhang2022graph}, e.g.,  updating node features ~\cite{wu2020joint}, modifying edges ~\cite{yang2021enhanced}. Another approach focuses on enhancing the robustness of GNN to the input graph structure, e.g., self-supervised learning based methods~\cite{gao2021rcd}, or adversarial learning based methods~\cite{sun2022adversarial}.

Among these approaches, graph refinements based on information bottleneck principles have garnered significant attention due to relatively-high performance and efficiency~\cite{wu2020graph,sun2022graph}. These principles mainly try to capture minimal yet sufficient information hidden in the graph structure while discarding information irrelevant to downstream tasks~\cite{wu2020graph}. Researchers have substantiated that these principles can help GNN effectively mitigate adversarial attacks~\cite{wang2023toward}. Yang et al. successfully preserve minimal yet sufficient information in social graphs for social recommendations~\cite{yang2024graph}. In this paper, we propose to minimize effects of edge uncertainty from the information bottleneck perspective. 
This process requires maximizing mutual information between a reliable graph and response logs, and minimizing mutual information between the reliable graph and the original graph. 

\section{The Proposed Model}
In this section, we first present important notations and task formulation, followed by the overall structure of our proposed \shortname. After that, we introduce each component of \shortname~in detail, and alternating training for \shortname. Finally, we discuss the time and space complexity of \shortname. 

\subsection{Task Formulation}
There are students {\small$S (|S|=M)$}, exercises {\small$E (|E|=N)$}, and  concepts {\small$K (|K|=T)$}.  
There are two types of relationships.  
First, students practice some exercises, forming response logs. {\small$r_{se}$} is student $s$'s response log to exercise $e$. 
We use {\small$R=\{(s,e,r_{se})\}$} to denote triplets of students, exercises and response logs. 
If student $s$ answers exercise $e$ correctly, {\small$r_{se}=1$}. Otherwise, {\small$r_{se}=0$}. 
In this paper, we also treat response logs as a bipartite graph {\small$\mathcal{G}$}. 
Second, the relations between exercises and concepts are denoted by {\small$\mathbf{Q}=\{q_{ek}\}_{N\times T}$}. If exercise $e$ is related to concept $k$, {\small$q_{ek}=1$}; otherwise, {\small$q_{ek}=0$}. 
Usually, {\small$\mathbf{Q}$} is pre-defined by experts. 
{\small$\mathbf{A} =[\mathbf{a}_1,...,\mathbf{a}_s,...,\mathbf{a}_M]^\top \in \mathbb{R}^{M\times T}$} represents student abilities (i.e., proficiency levels on all concepts), where {\small$\mathbf{a}_s$} denotes student $s$'s ability.  
Each dimension of {\small$\mathbf{a}_s=[{a}_{s1},...,{a}_{sk},...,{a}_{sT}]$} has independent meaning, e.g., {\small${a}_{sk}$} denotes student $s$'s  proficiency level on concept $k$. 
There are two exercise-side parameters, i.e., exercise difficulties {\small$\mathbf{D} =[\mathbf{d}_1,...,\mathbf{d}_e,...,\mathbf{d}_N]^\top \in \mathbb{R}^{N\times T}$} and discriminations {\small$\mathbf{h}^{disc} =[{h}_1^{disc},...,{h}_e^{disc},...,{h}_N^{disc}]^\top \in \mathbb{R}^{N\times 1}$}. 
Based on these notations, we formulate CD as follows, \newline
\textbf{Input:} The response logs {\small${R}$} and exercise-concept relations {\small$\mathbf{Q}$}. \newline
\textbf{Output:} A model to diagnose students' abilities (proficiency levels on all concepts) through response log prediction. 

\subsection{Overall Structure of \shortname}
\begin{figure}[t]
    \includegraphics[width=0.45\textwidth]{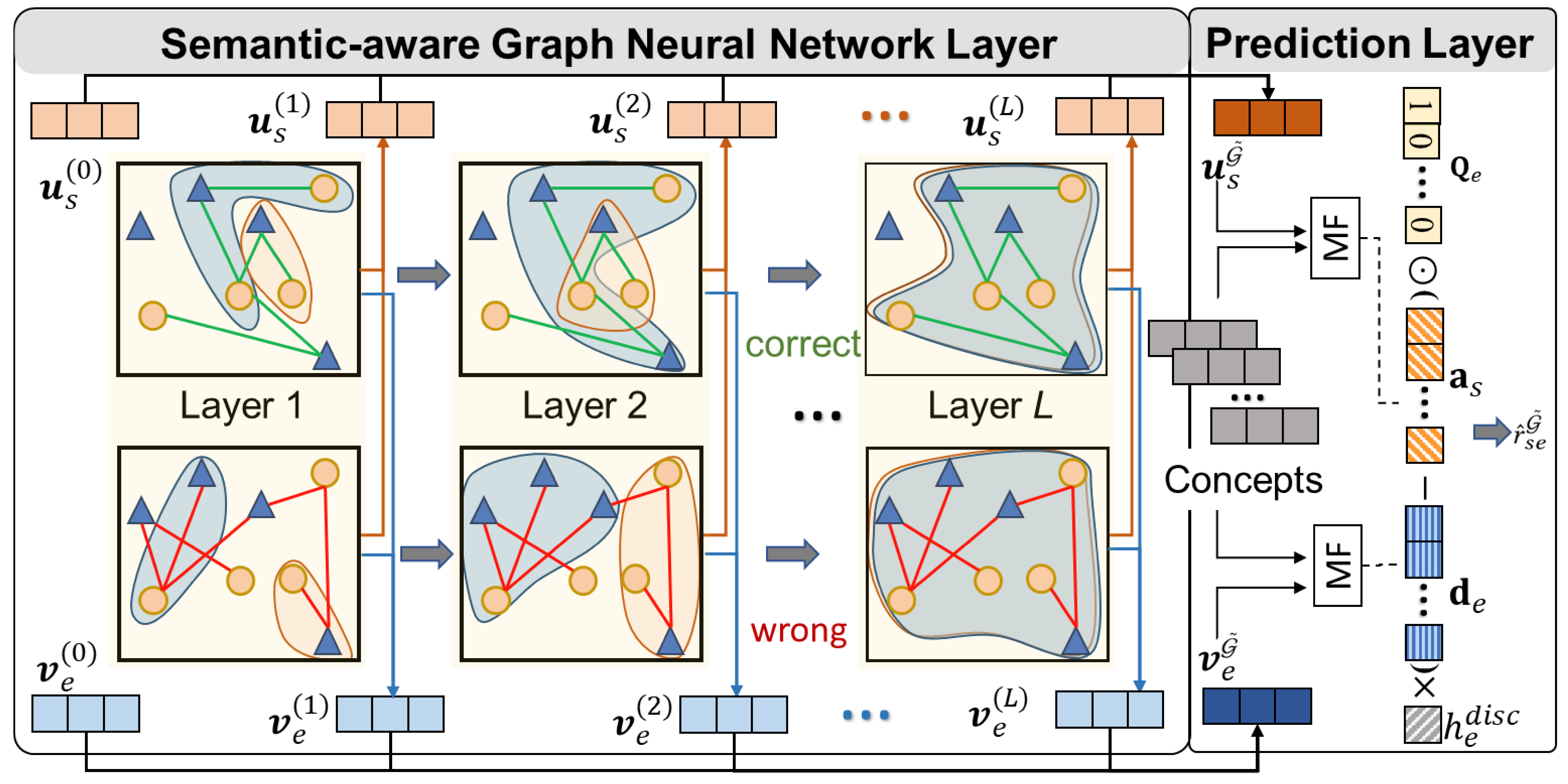}
    \caption{ Semantic-aware GNN (S-GNN) based CD model.}
    \label{fig:model}
\end{figure}

Compared to the backbone KaNCD~\cite{wang2022neuralcd}, there are two additional components in \shortname.
Figure \ref{fig:model} describes a semantic-aware GNN layer based CD model. S-GNN focuses on how to make full use of a reliable student-exercise graph for CD. After that, Figure \ref{fig:model2} describes the informative edge differentiation (IE-Diff) layer, which is designed for obtaining the reliable graph based on information bottleneck principles~\cite{wang2021revisiting}. 

These two layers are complementary to each other. On one hand, the IE-Diff layer relies on student and exercise latent embeddings from the S-GNN layer. On the other hand, the S-GNN layer relies on the reliable graph provided by the IE-Diff layer.  Hence, we adopt alternately training these two layers~\cite{zibetti2022alternating,zhu2024collaborative}. 
In the following parts of this section, we will discuss these two components.

\subsection{Semantic-aware GNN Based CD Model}
\label{sec:model1}
\subsubsection{Embedding Module and Semantic-aware GNN (S-GNN) Layer}
Figure \ref{fig:model} describes incorporating the S-GNN layer into a CD model. 
We adopt KaNCD as our backbone  due to its relatively-high performance and easy-to-implementation~\cite{wang2022neuralcd}. 
There are three latent embeddings in KaNCD, i.e.,  {\small$\mathbf{U} = [ \mathbf{u}_1,..., \mathbf{u}_s,..., \mathbf{u}_M]^\top \in \mathbb{R}^{M \times Z}$} for students, {\small$\mathbf{V}=[\mathbf{v}_1,..., \mathbf{v}_e,...,\mathbf{v}_N]^\top \in \mathbb{R}^{N \times Z}$} for exercises  and {\small$\mathbf{O}=[\mathbf{o}_1,..., \mathbf{o}_k$,} {\small$..., \mathbf{o}_T]^\top \in \mathbb{R}^{T \times Z}$} for knowledge concepts. {\small$Z$} denotes the dimension of latent embeddings. 

Inspired by recent success on graph learning~\cite{gao2021rcd,li2022hiercdf,hu2020heterogeneous}, we focus on utilizing the student-exercise graph to enhance student and exercise latent embeddings. 
There are two types of edges with fundamentally different semantic meanings in this graph: correct answers and incorrect answers. Previous studies ignored this edge heterogeneity, which poses a risk of confounding semantic information in GNN-based CD models~\cite{gao2021rcd,wang2023self}. To capture heterogeneity, we propose a S-GNN layer.
Here, we suppose that we have already obtained the reliable graph {\small$\tilde{\mathcal{G}}$}. We split {\small$\tilde{\mathcal{G}}$} into subgraphs based on semantics ({\small$\tilde{\mathcal{G}} = \tilde{\mathcal{G}}_1 \cup \tilde{\mathcal{G}}_0$}). 
{\small$\tilde{\mathcal{G}}_1$}, {\small$\tilde{\mathcal{G}}_0$} denote subgraphs corresponding to correct/incorrect response logs, respectively. 
We treat latent embeddings of students and exercises in KaNCD as local embeddings, e.g., {\small$\mathbf{v}_e = \mathbf{v}_e^{(0)}$}, {\small$\mathbf{u}_s = \mathbf{u}_s^{(0)}$}. 
Edge-specific messages {\small$\mu$} from exercise $e$ to student $s$ can be divided into two parts: 
\begin{small}
    \begin{equation}
        \label{eq:mu}
        \mu_{e\rightarrow s,\tilde{\mathcal{G}}_1}^{(l)} = \frac{1}{\sqrt{|\mathcal{N}_{s,\tilde{\mathcal{G}}_1}||\mathcal{N}_{e,\tilde{\mathcal{G}}_1}|}}\mathbf{v}_{e}^{(l-1)}, \mu_{e\rightarrow s,\tilde{\mathcal{G}}_0}^{(l)} = \frac{1}{\sqrt{|\mathcal{N}_{s,\tilde{\mathcal{G}}_0}||\mathcal{N}_{e,\tilde{\mathcal{G}}_0}|}}\mathbf{v}_{e}^{(l-1)},
    \end{equation}
\end{small}
where  {\small$\mu_{e\rightarrow s,\tilde{\mathcal{G}}_1}^{(l)}$} and {\small$\mu_{e\rightarrow s,\tilde{\mathcal{G}}_0}^{(l)}$} denote message passed in subgraph {\small$\tilde{\mathcal{G}}_1$} and {\small$\tilde{\mathcal{G}}_0$} at layer $l$, respectively. 
{\small$\mathcal{N}_{s,\tilde{\mathcal{G}}_1},\mathcal{N}_{e,\tilde{\mathcal{G}}_1}$} denote the neighbor sets of student $s$ and exercise $e$ in subgraph {\small$\tilde{\mathcal{G}}_1$}. 
Similarly, {\small$\mathcal{N}_{s,\tilde{\mathcal{G}}_0},\mathcal{N}_{e,\tilde{\mathcal{G}}_0}$} denote the neighbor sets in subgraph {\small$\tilde{\mathcal{G}}_0$}. 
{\small$\mathbf{v}_{e}^{(l-1)}$} denotes exercise $e$'s difficulty at layer {\small$(l-1)$}. 
Messages from student $s$ to exercise $e$ can be processed analogously, formulated as: 
\begin{small}
    \begin{equation}
    \label{eq:mu2}
        \mu_{s\rightarrow e,\tilde{\mathcal{G}}_1}^{(l)} = \frac{1}{\sqrt{|\mathcal{N}_{s,\tilde{\mathcal{G}}_1}||\mathcal{N}_{e,\tilde{\mathcal{G}}_1}|}}\mathbf{u}_{s}^{(l-1)}, 
        \mu_{s\rightarrow e,\tilde{\mathcal{G}}_0}^{(l)} = \frac{1}{\sqrt{|\mathcal{N}_{s,\tilde{\mathcal{G}}_0}||\mathcal{N}_{e,\tilde{\mathcal{G}}_0}|}}\mathbf{u}_{s}^{(l-1)},
    \end{equation}
\end{small}
where {\small$\mathbf{u}_{s}^{(l-1)}$} denotes student $s$'s latent embedding at layer {\small$(l-1)$}. After message passing, we can obtain incoming messages at every node from two types of edges. In message aggregation, for each node, we first sum incoming messages over all neighbors under a specific edge type. Subsequently, we accumulate the messages from all edge-types into vector representations at the $l$-th layer:
\begin{small}
    \begin{equation}
    \begin{aligned}
    \label{eq:aggregation}
        \mathbf{u}_s^{(l)} = 
        \sum_{e} \mu_{e\rightarrow s, \tilde{\mathcal{G}}_1} + \sum_{e} \mu_{e\rightarrow s, \tilde{\mathcal{G}}_0}, 
        ~\mathbf{v}_e^{(l)} = 
        \sum_{s} \mu_{s\rightarrow e, \tilde{\mathcal{G}}_1}+\sum_{s} \mu_{s\rightarrow e, \tilde{\mathcal{G}}_0}. 
    \end{aligned}
    \end{equation}
\end{small}


We consider the following read out functions for final embeddings: {\small$\mathbf{u}_s^{\tilde{\mathcal{G}}} = \sum_{l=0}^{L}\mathbf{u}_s^{(l)}, \mathbf{v}_e^{\tilde{\mathcal{G}}} = \sum_{l=0}^{L}\mathbf{v}_e^{(l)}$}. The superscripts of  {\small$\mathbf{u}_s^{\tilde{\mathcal{G}}}$} and  {\small$\mathbf{v}_e^{\tilde{\mathcal{G}}}$} indicate that these representations are derived from graph {\small$\tilde{\mathcal{G}}$}.

\subsubsection{Matrix Factorization and Prediction Layer}
\label{sec:model3}
As shown in Figure \ref{fig:model}, we combine latent embeddings to obtain students' proficiency levels and exercise difficulties via matrix factorization following KaNCD~\cite{wang2022neuralcd}. In this paper, we adopt a simple yet effective probabilistic matrix factorization, which is widely adopted in recommender systems~\cite{mnih2007probabilistic,wu2023causality,gao2023cirs}. The process is as follows: 
\begin{small}
    \begin{equation}
    \label{eq:pmf}
        a_{sk} = \sigma(<\mathbf{u}_s^{\tilde{\mathcal{G}}},\mathbf{o}_k>), ~{d}_{ek} = \sigma(<\mathbf{v}_e^{\tilde{\mathcal{G}}},\mathbf{o}_k>), 
    \end{equation}
\end{small}
where {\small$<,>$} denotes the inner dot between two embeddings. {\small$\sigma$} denotes the sigmoid activation function. By repeating Eq.(\ref{eq:pmf}) on all concepts, we can obtain student's proficiency levels and exercise difficulties, i.e., {\small$\mathbf{a}_s$} and {\small$\mathbf{d}_e$}. Following~\cite{FCS2024shao}, we allow adding bias terms during the above process. 
The connections between {\small$\mathbf{a}_s$}, {\small$\mathbf{d}_e$} and response logs can be represented as that a student is more likely to answer an exercise correctly when the student's ability exceeds the exercise difficulty on corresponding knowledge concepts. Therefore, the diagnostic function can be formulated as:
\begin{small}
\begin{equation}
\label{eq:diagnostic_function}
\begin{aligned}
    \mathbf{p}_{se}^{\tilde{\mathcal{G}}}  = \mathbf{Q}_e \odot (\mathbf{a}_s-\mathbf{d}_e) \times {h}_e^{disc}, ~\hat{r}_{se}^{\tilde{\mathcal{G}}}  =\sigma(MLPs(\mathbf{p}_{se}^{\tilde{\mathcal{G}}})),
\end{aligned}
 \end{equation}
 \end{small}
where {\small$\mathbf{Q}_e$} denotes the relations between concepts and exercise $e$, and {\small$\odot$} denotes the element-wise product. {\small$\mathbf{p}_{se}^{\tilde{\mathcal{G}}}$} denotes the hidden representation between student $s$ and exercise $e$, and {\small$\hat{r}_{se}^{\tilde{\mathcal{G}}}$} denotes the predicted student $s$'s response log on exercise $e$ based on graph {\small${\tilde{\mathcal{G}}}$}. Finally, we use the classic BCE loss, formulated as: 
\begin{small}
\begin{equation}
\label{eq:BCE_loss}
    \min_{\theta}\mathcal{L}_{BCE} = -\sum_{(s,e,r_{se}) \in R} (r_{se}\log(\hat{r}_{se}^{\tilde{\mathcal{G}}}) + (1-r_{se})\log(1-\hat{r}_{se}^{\tilde{\mathcal{G}}})). 
 \end{equation}
 \end{small}
 where {\small$\theta$} denotes all trainable parameters of graph-based CD models, i.e.,  trainable parameters in Section \ref{sec:model1}. The remaining problem relies on how to obtain the reliable graph {\small$\tilde{\mathcal{G}}$}. 

\begin{figure}[t]
    \includegraphics[width=0.45\textwidth]{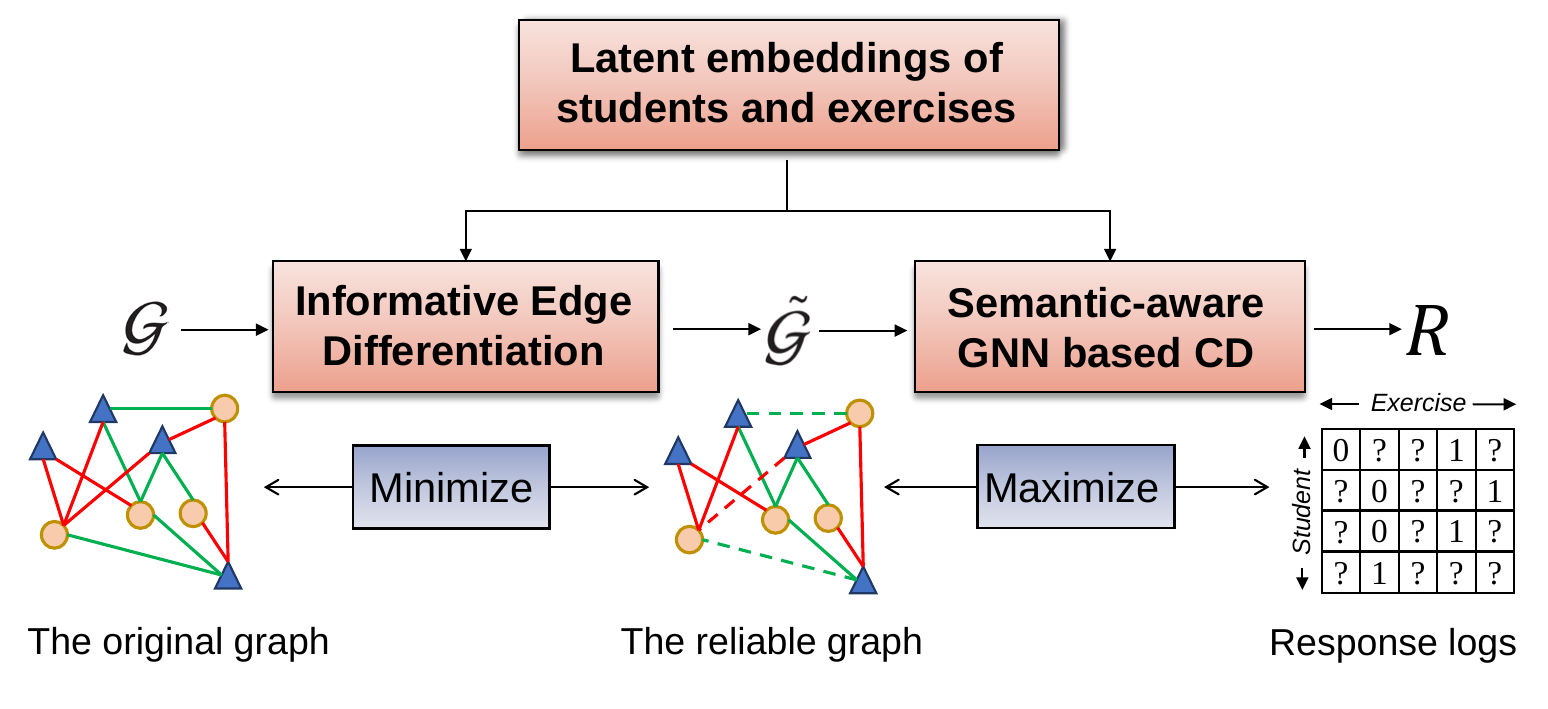}
    \caption{Informative Edge Differentiation (IE-Diff) Layer. }
    \label{fig:model2}
\end{figure}

\subsection{Informative Edge Differentiation Layer}
\label{sec:model2}
In this part, we aim to build a reliable student-exercise graph  {\small$\tilde{\mathcal{G}}$} based on the original graph {\small${\mathcal{G}}$}. 
To improve diagnostic performance, the reliable graph {\small$\tilde{\mathcal{G}}$} should minimize the effects of uncertain edges. 
To this end, we propose an Informative Edge Differentiation (IE-Diff) layer to obtain this reliable graph.   
Our intuition is from a powerful information bottleneck   principle~\cite{tishby2015deep,wang2021revisiting}, which focuses on extracting relevant information from data while simultaneously discarding irrelevant details. Formally, given input data {\small$X$}, hidden representations {\small$H$}, and downstream task labels {\small$Y$} (chain form: {\small$<X\rightarrow H\rightarrow Y>$}), the information bottleneck principle suggests that optimal representations should keep minimal yet sufficient information for downstream tasks~\cite{saxe2019information,yang2024graph}, formulated as:
\begin{small}
\begin{equation}
\label{eq:id0}
\begin{aligned}
    H^*={argmax}_{H} I(Y;H)-\beta I(X;H), 
\end{aligned}
 \end{equation}
 \end{small}
where {\small$I(Y;H)$} and {\small$I(X;H)$} denote the mutual information between corresponding variables, and {\small$\beta$} denotes the balancing parameter between two parts. As shown in Figure \ref{fig:model2}, we adopt the information bottleneck principle in  reducing effects of uncertain edges in the student-exercise bipartite graph, formulated as: 
\begin{small}
\begin{equation}
\label{eq:ib}
\begin{aligned}
\max_{\phi}  {I}({R};\mathbf{U},\mathbf{V},\tilde{\mathcal{G}}) - \beta {I}(\tilde{\mathcal{G}},\mathcal{G}), 
\end{aligned}
 \end{equation}
 \end{small}
 where {\small$R$} denotes response logs, {\small$\mathbf{U},\mathbf{V}$} denote latent embeddings of students and exercises. {\small$\mathcal{G}$} and {\small$\tilde{\mathcal{G}}$} denote the original bipartite graph and the reliable graph, respectively. 
 {\small$\phi$} denotes parameters of the reliable graph {\small$\tilde{\mathcal{G}}$}. 
 Eq.(\ref{eq:ib}) requires maximizing {\small${I}({R};\mathbf{A},\mathbf{D},\tilde{\mathcal{G}})$} and minimizing {\small${I}(\tilde{\mathcal{G}},\mathcal{G})$}. 
 One obvious challenge in realization Eq.(\ref{eq:ib}) lies in how to parameterize {\small$\tilde{\mathcal{G}}$}. 

\subsubsection{Parameterization of the reliable graph {$\tilde{\mathcal{G}}$}} 
In Eq.(\ref{eq:ib}), {\small$\tilde{\mathcal{G}}$} should be learnable. Typically, a student-exercise graph  can be expressed as a binary adjacency matrix, while this matrix itself can not be optimized. 
Therefore, a key challenge lies in the parameterization of {\small$\tilde{\mathcal{G}}$}. 
In this paper, we formulate the edge parameterization as a graph edge dropout problem~\cite{yang2024graph}. The detailed form is {\small$\tilde{g}_{se} = {g}_{se}\odot \rho_{se}$}, where  {\small${g}_{se}$} and {\small$\tilde{g}_{se}$} denote the edge between student $s$ and exercise $e$ in the original graph {\small${\mathcal{G}}$} and reliable graph {\small$\tilde{\mathcal{G}}$}, respectively. {\small$\odot$} denotes the element-wise product. {\small${\rho}_{se}$}  describes the certainty of edge {\small${g}_{se}$}. The more uncertain the edge is, the higher the probability of being dropped. 
We model edge certainty for different subgraphs. 
Depending on whether edge {\small$g_{se}$} corresponds to a correct response log or an incorrect response log, we choose {\small$\rho_{se}^1$} and {\small$\rho_{se}^0$} to model its certainty, respectively. {\small$\rho_{se}^1$} and {\small$\rho_{se}^0$} together form the overall {\small$\rho_{se}$}. 
Then, we model them as {\small${\rho}_{se}^1 \sim Bern(w_{se}^1) $} and {\small${\rho}_{se}^0 \sim Bern(w_{se}^1)$}. 
{\small$Bern(w_{se}^1)$} and {\small$Bern(w_{se}^0)$} represent Bernoulli distributions with parameter {\small$w_{se}^1$} and {\small$w_{se}^0$}, respectively. Then, we propose to parameterize {\small$w_{se}^1$} and {\small$w_{se}^0$} with considering different semantics as:
 \begin{small}
\begin{equation}
\label{eq:bern0}
\begin{aligned} 
{w}_{se}^0 =\mathbf{W}^0([\mathbf{u}_s;\mathbf{v}_e]), ~{w}_{se}^1 =\mathbf{W}^1([\mathbf{u}_s;\mathbf{v}_e]),
\end{aligned}
 \end{equation}
 \end{small}
where {\small$\mathbf{W}^0,\mathbf{W}^1$} denotes corresponding trainable weight matrices. 
{\small$[;]$} denotes concatenation of two embeddings. After obtaining {\small$w_{se}^0$} and {\small$w_{se}^1$}, we focus on 
making {\small$Bern(w_{se}^0)$} and {\small$Bern(w_{se}^1)$} differentiable. We adopt the popular concrete relaxation method~\cite{jang2016categorical}: 
 \begin{small}
\begin{equation}
\label{eq:bern}
\begin{aligned}
    Bern(w_{se}^0) = \sigma (\log(\delta/(1-\delta)+w_{se}^0)/t), \\
     Bern(w_{se}^1) = \sigma (\log(\delta/(1-\delta)+w_{se}^1)/t), 
\end{aligned}
 \end{equation}
 \end{small}
 where $t$ is the temperature parameter and we set $t=0.2$ in our experiments. {\small$\delta$} denotes random variable sampled from $U(0,1)$.

\subsubsection{Maximization of {\small${I}({R};\mathbf{U},\mathbf{V},\tilde{\mathcal{G}})$}} In this part, we focus on how to handle the first term in Eq.(\ref{eq:ib}). {\small${I}({R};\mathbf{U},\mathbf{V},\tilde{\mathcal{G}})$} requires that the reliable graph {\small$\tilde{\mathcal{G}}$} should satisfy the response log prediction. Here, we derive the lower bound of {\small${I}({R};\mathbf{U},\mathbf{V},\tilde{\mathcal{G}})$} for its maximization. 
\begin{small}
\begin{equation}
\label{eq:infer}
\begin{aligned}
    {I}({R};\mathbf{U},\mathbf{V},\tilde{\mathcal{G}}) &= H({R}) - H({R}|\mathbf{U},\mathbf{V},\tilde{\mathcal{G}}) \\
     & \stackrel{(a)}{\geq}\sum_{s=1}^{M} \sum_{e=1}^{N} \sum_{r=0}^{1} p(r_{se},\mathbf{u}_s,\mathbf{v}_e, \mathcal{G}) \log p(r_{se}|\mathbf{u}_s,\mathbf{v}_e, \tilde{\mathcal{G}}) \\ 
    & \stackrel{(b)}{\geq} \sum_{(s,e,r_{se}) \in R} \log p(r_{se}|\mathbf{u}_s,\mathbf{v}_e, \tilde{\mathcal{G}}) \\
    & \stackrel{(c)}{=} \sum_{(s,e,r_{se}) \in R} \log p(r_{se}|\hat{r}_{se}) \\ 
    & \stackrel{(d)}{=}\sum_{(s,e,r_{se}) \in R} \log [{\hat{r}_{se}}^{{r}_{se}} (1-{\hat{r}_{se}})^{(1-{r}_{se})}] \\ 
    & \stackrel{(e)}{=} \sum_{(s,e,r_{se}) \in R} (r_{se}\log(\hat{r}_{se}) + (1-r_{se})\log(1-\hat{r}_{se})).   
\end{aligned}
 \end{equation}
 \end{small}
 Here, {\small$\hat{r}_{se}$} is short for  {\small$\hat{r}_{se}^{\tilde{\mathcal{G}}}$}. 
 The reason for each derivation step is as follows: (a) is the non-negative property of entropy; (b) is that {\small$p(r_{se},\mathbf{u}_s,\mathbf{v}_e, \mathcal{G}) \leq 1$} and {\small$log p(r_{se}|\mathbf{u}_s,\mathbf{v}_e, \tilde{\mathcal{G}}) < 0$}; (c) is that we can obtain {\small$\hat{r}_{se}$} given students and exercises' parameters (this process is shown in Figure \ref{fig:model}); (d) represents the form in the context of maximum likelihood probability; (e) is the property of {\small$log$} function. 
The lower bound of {\small${I}({R};\mathbf{A},\mathbf{D},\tilde{\mathcal{G}})$} is the negative of a Binary Cross-Entropy (BCE) function (a widely-adopted loss in CD models~\cite{wang2022neuralcd,wang2023self}). Therefore, maximizing the mutual information is equivalent to minimizing the BCE loss in Eq.(\ref{eq:BCE_loss}). 

\subsubsection{Minimization of {\small${I}(\tilde{\mathcal{G}},\mathcal{G})$}}
Next, we focus on how to minimize the second term in Eq.(\ref{eq:ib}). Estimating the upper bound of mutual information is difficult. Some methods employ variational techniques for this estimation~\cite{alemi2016deep,bang2021explaining,abdelaleem2023deep}, however, they heavily depend on prior assumptions. In this paper, we  adopt the Hilbert-Schmidt Independence Criterion (\textbf{HSIC}~\cite{gretton2005measuring,ma2020hsic}) for mutual information minimization. 
{\small${HSIC}(X, Y)$} measures dependence between two variables {\small$X$} and {\small$Y$}. Generally, if {\small${HSIC}(X, Y)$} is closer to 0, these variables are more independent from each other. In detail, {\small${HSIC}(X, Y)$} is based on covariance operators in the Reproducing Kernel Hilbert Space~\cite{berlinet2011reproducing}, formulated as: 
 \begin{small}
\begin{equation}
\label{eq:HSIC0}
\begin{aligned}
 {HSIC}(X, Y) = & \mathbb{E}_{X,X',Y,Y'}[\mathcal{K}_X(X, X')\mathcal{K}_Y(Y, Y')] \\
 + &\mathbb{E}_{X,X'}[\mathcal{K}_X(X, X')] \mathbb{E}_{Y,Y'}[\mathcal{K}_Y(Y, Y')]  \\
 - & 2 \mathbb{E}_{XY}[\mathbb{E}_{X'}[\mathcal{K}_X(X, X')] \mathbb{E}_{Y'}[\mathcal{K}_Y(Y, Y')]], 
\end{aligned}
 \end{equation}
 \end{small}
where {\small$X'$} and {\small$Y'$} are two independent copies of {\small$X$} and {\small$Y$}. {\small$\mathcal{K}_X$} and {\small$\mathcal{K}_Y$} denote the kernel function for two variables. Given sufficient   instances {\small$\{x_i,y_i\}_{i=1}^n$} in the batched training samples, estimation of {\small${HSIC}(X, Y)$} can be realized as {\small${HSIC}(X, Y)=(n-1)^{-2}Tr(\mathcal{K}_X{J}\mathcal{K}_Y{J})$}, where 
{\small$Tr(\cdot)$} denotes the trace of a matrix. {\small$J=\mathbf{I}-\mathbf{1}\mathbf{1}^T/n$} denotes the centering matrix, where {\small$\mathbf{I}$} denotes the Identity matrix, {\small$n$} denotes the number of batched instances~\cite{wang2021learning}. The inputs of {\small$\mathcal{K}_X$} and {\small$\mathcal{K}_Y$} are two instances from batched samples, i.e., {\small$\mathcal{K}_X(x_i,x_j)$} and {\small$\mathcal{K}_Y(y_i,y_j)$}. In this paper, we adopt a widely-adopted Radial Basis Function (RBF) for {\small$\mathcal{K}_X$}~\cite{vert2004primer}: 
 \begin{small}
\begin{equation}
\label{eq:RBF}
\begin{aligned}
\mathcal{K}_X(x_i,x_j) = exp(-\frac{||x_i-x_j||^2}{2\alpha}), 
\end{aligned}
 \end{equation}
 \end{small}
where {\small$\alpha$} denotes a constant that controls the sharpness of {\small$\mathcal{K}_X$}.  
Following~\cite{yang2024graph}, we set {\small$\alpha$} to 0.2 in experiments.
We also adopt the same RBF kernel function for {\small$\mathcal{K}(y_i,y_j)$}. In this paper, we aim to minimize mutual information between two graphs. Here, we utilize the minimization of {\small$HSIC(\tilde{\mathcal{G}},\mathcal{G})$} to replace direct minimization of {\small${I}(\tilde{\mathcal{G}},\mathcal{G})$}. 
As graphs are non-Euclidean, we adopt Monte Carlo sampling on node representations for estimation~\cite{yang2024graph}: {\small$HSIC(\tilde{\mathcal{G}},\mathcal{G})) = HSIC({\mathbf{U}^{\tilde{\mathcal{G}}}},\mathbf{U}^{{\mathcal{G}}})+HSIC({\mathbf{V}^{\tilde{\mathcal{G}}}},\mathbf{V}^{{\mathcal{G}}})$}, 
where {\small${\mathbf{U}^{\tilde{\mathcal{G}}}}$} and {\small${\mathbf{V}^{\tilde{\mathcal{G}}}}$} denote final embeddings based on the reliable graph {\small$\tilde{\mathcal{G}}$}.  {\small${\mathbf{U}^{{\mathcal{G}}}}$} and {\small${\mathbf{V}^{{\mathcal{G}}}}$} denote embeddings based on the original graph {\small${\mathcal{G}}$}. In our experiments, we take students and exercises in each batch into consideration.  In summary, we realize minimizing mutual information via the following HSIC-based loss function: 
 \begin{small}
\begin{equation}
\label{eq:HSIC}
\begin{aligned}
 {L}_{HSIC} = HSIC({\mathbf{U}^{\tilde{\mathcal{G}}}},\mathbf{U}^{{\mathcal{G}}})+HSIC({\mathbf{V}^{\tilde{\mathcal{G}}}},\mathbf{V}^{{\mathcal{G}}}). 
\end{aligned}
 \end{equation}
 \end{small}
We combine Eq.(\ref{eq:BCE_loss}) and Eq.(\ref{eq:HSIC}) for the final loss, which can be formulated as: 
 \begin{small}
\begin{equation}
\label{eq:final_loss}
\begin{aligned}
 \min_{\phi}\mathcal{L}_{all} = \mathcal{L}_{BCE} + \beta \mathcal{L}_{HSIC},  
\end{aligned}
 \end{equation}
 \end{small}
where {\small$\mathcal{L}_{all}$} is used to update parameters of IE-Diff ({\small$\phi$}). 

\subsection{Alternating Training Strategy}
A straightforward idea is to adopt the paradigm of multi-task learning and use Eq.(\ref{eq:final_loss}) to optimize overall parameters of \shortname~  together ({\small$\theta$} and {\small$\phi$})~\cite{yang2024graph}. Considering the interdependence between the S-GNN based CD model and the IE-Diff layer, we adopt alternating training ~\cite{zibetti2022alternating,zhu2024collaborative}. 
Specifically, we choose to pre-train {\small$\theta$} based on the original graph for a few epochs as a start. In practice, we allow for 5 additional pre-training epochs on all three datasets. Subsequently, we alternately perform two steps: fix {\small$\theta$} to optimize {\small$\phi$} (Eq.(\ref{eq:final_loss})), then fix {\small$\phi$} to optimize {\small$\theta$} (Eq.(\ref{eq:BCE_loss})) until convergence. To better highlight the advantages, we also compare different strategies in Section \ref{sec:strategies}.

\subsection{Model Discussions}
\subsubsection{Additional Time Complexity}
Compared to the backbone model KaNCD, \shortname~ first involves with  GNN. Specifically, we utilize sparse tensors to represent adjacency matrices. The Update of student/exercise latent embeddings relies on a sparse tensor multiplication, and the time complexity is $O(nnz*T)$, where $T$ denotes the number of concepts and $nnz$ represents the number of non-zero elements in the sparse matrix. Second, we introduce an informative edge differentiation layer. 
As shown in Eq.(\ref{eq:bern}), this layer maps local embeddings to certainty for all edges. Note that, this process does not involve with GNNs, therefore, the time complexity caused by IE-Diff is much lower than S-GNN. 
Overall, the additional time complexity of \shortname~ is low.

\subsubsection{Additional Space Complexity}
 \shortname~ has two parts of additional parameters compared to KaNCD. One part lies in adjacency matrices corresponding to subgraphs, whose space complexity of sparse matrices is $O(nnz)$. 
The other part is informative edge differentiation, which applies additional MLPs to calculate edge certainty. The additional MLPs are shared across all edges, therefore, space complexity of this part can be ignored. 
Overall, the additional space complexity of \shortname~ is low.

\subsubsection{Training Procedures of \shortname}
\label{sec:procedure}
For readability, we present the training procedures of \shortname~ in Algorithm 1. The process is divided into updating the parameters of the graph structure and the parameters of the CD model sequentially within each epoch. 
\begin{algorithm}[H]
\caption{Detailed training procedures of \shortname.} 
\label{alg:propensity}
\begin{algorithmic}[1] 
\renewcommand{\algorithmicrequire}{\textbf{Require:}}
\REQUIRE 
Triplets of students, exercises and response log $R$, student-exercise bipartite graph $\mathcal{G}$. 
\renewcommand{\algorithmicensure}{ \textbf{Ensure:}}
\ENSURE 
\STATE Initialize parameters of graph-based CD ({$\theta$}) and parameters of edge differentiation layer ({$\phi$}). 
\STATE Pre-train $\theta$ based on the original graph {\small$\mathcal{G}$} for several epochs. 
\REPEAT
\STATE Fix {$\theta$} and unfix $\phi$; 
\FOR{each sample $(s,e,r_{se})$ in the training data}
\STATE Obtain the reliable graph {\small$\tilde{\mathcal{G}}$} (Eq.(\ref{eq:bern}));
\STATE Update ${\mathbf{U}^{\tilde{\mathcal{G}}}},{\mathbf{V}^{\tilde{\mathcal{G}}}}$ according to {\small$\tilde{\mathcal{G}}$} (Eq.(\ref{eq:aggregation})); 
\STATE Update ${\mathbf{U}^{{\mathcal{G}}}},{\mathbf{V}^{{\mathcal{G}}}}$ according to {\small${\mathcal{G}}$} (Eq.(\ref{eq:aggregation})); 
\STATE Obtain student abilities and exercise difficulties based on matrix factorization (Eq.(\ref{eq:pmf})); 
\STATE Obtain response log prediction {\small$\hat{r}_{se}^{\tilde{\mathcal{G}}}$} (Eq.(\ref{eq:diagnostic_function})); 
\STATE Calculate HSIC loss (Eq.\ref{eq:HSIC});
\STATE Calculate BCE loss (Eq.(\ref{eq:BCE_loss})); 
\STATE Calculate overall loss (Eq.(\ref{eq:final_loss})); 
\ENDFOR 
\STATE Minimize Eq.(\ref{eq:final_loss})  to update $\phi$,

\STATE Fix $\phi$ and unfix $\theta$; 
\FOR{each sample $(s,e,r_{se})$ in the training data}
\STATE Obtain the reliable graph {\small$\tilde{\mathcal{G}}$} (Eq.(\ref{eq:bern}));
\STATE Update ${\mathbf{U}^{\tilde{\mathcal{G}}}},{\mathbf{V}^{\tilde{\mathcal{G}}}}$ according to {\small$\tilde{\mathcal{G}}$} (Eq.(\ref{eq:aggregation})); 
\STATE Obtain student abilities and exercise difficulties based on matrix factorization (Eq.(\ref{eq:pmf})); 
\STATE Obtain response log prediction {\small$\hat{r}_{se}^{\tilde{\mathcal{G}}}$} (Eq.(\ref{eq:diagnostic_function})); 
\STATE Calculate BCE loss (Eq.(\ref{eq:BCE_loss}));  
\ENDFOR 
\STATE Minimize BCE loss to update $\theta$,
\UNTIL{Convergence.} 
\end{algorithmic}
\end{algorithm}

\section{Experiments}

In this section, we try to answer these Research Questions~(RQ):
\begin{itemize}[leftmargin=0.9cm]
\item [\textbf{RQ1:}] Does \shortname~have consistently superior performance on three datasets? (Section \ref{sec:overallperformance})
\item [\textbf{RQ2:}] Does \shortname~really detect uncertain edges?  (Section \ref{sec:detect})
\item [\textbf{RQ3:}] Are all components in \shortname~important? (Section \ref{sec:ablation})
\item [\textbf{RQ4:}] What are the impacts of different hyperparameters on \shortname? Can \shortname~consistently achieve better diagnostic performance than the second-best CD model? (Section \ref{sec:hyper})
\item [\textbf{RQ5:}] What are the impacts of different training strategies on \shortname? (Section \ref{sec:strategies})

\end{itemize}

\subsection{Experimental Settings}
\subsubsection{Datasets.} In the experimental part, we choose three real-world datasets, i.e., {ASSIST} dataset (ASSISTments 2009-2010 ”skill builder”)\footnote{https://sites.google.com/site/assistmentsdata/feng2009},  Junyi dataset (Junyi Academy Math
Practicing Log)\footnote{https://pslcdatashop.web.cmu.edu/DatasetInfo?datasetId=1198}, and MOOC-Radar dataset\footnote{https://github.com/THU-KEG/MOOC-Radar}.
{ASSIST} is a publicly available dataset collected by the ASSISTments online tutoring systems~\cite{feng2009addressing}. It contains a wealth of student-exercise response logs and expert-labeled exercise-concept relations $\mathbf{Q}$. 
Junyi dataset is an open dataset collected by an e-learning website, Junyi Academy. 
The unique characteristic of the Junyi dataset lies in one-to-one correspondence between knowledge concepts and exercises. 
We adopt the same pre-processing method as ~\cite{li2022hiercdf}.  
{MOOC-Radar} is a recently collected dataset from students' learning records in MOOCs~\cite{yu2023moocradar}, which contains abundant response logs. 
We record the detailed statistics in Table \ref{tab:datasets}. 
Finally, we conduct five-fold cross-validation for all models. 
Specifically, we randomly split all student-exercise response logs into 5 parts. Each part will be treated as the testing set in turn, and then we split the remaining 4 parts into the training set and the validation set with the ratio of 7:1. 

\begin{table}[t]
    \centering
    \caption{The detailed statistics of three datasets. } 
    \label{tab:datasets}
    \scalebox{0.95}{
    \begin{tabular}{cccc}
    \hline
        Dataset & {ASSIST} & {Junyi}& {MOOC-Radar}  \\ \hline
         \#Students & 2,493 & 10,000 & 14,224\\ 
         \#Exercises & 17,746 & 734& 2,513\\ 
         \#Knowledge concepts & 123 & 734&580\\ 
         \#Response logs & 267,415  & 408,057&898,933\\ 
         \#Response logs per student & 107.266  & 40.8&63.198 \\ 
         \#Concepts per exercise & 1.192 & 1&1\\
         \#Sparsity in response logs & 99.396\% & 94.441\% &97.485\%\\ 
          \hline
    \end{tabular}}
\end{table}

\subsubsection{Metrics.}
Similar to previous studies~\cite{gao2021rcd,li2022hiercdf,wang2020neural,wang2023self}, we evaluate the prediction performance of response logs. Two commonly used accuracy metrics have been adopted for performance evaluation: Accuracy (ACC), and Area Under the Curve (AUC).  
Additionally, we employ a widely-used metric called Degree of Agreement (DOA)~\cite{wu19edustudio, wang2022neuralcd}. DOA measures the consistency between the predicted proficiency levels and the observed patterns in response logs. 
Note that, we evaluate DOA on the testing set. 

\subsubsection{Baselines.}
We select the following cutting-edge and high-performing CD models as our baselines: 
\begin{itemize}[leftmargin=0.5cm]
\item \textbf{DINA}~\cite{de2009dina}. It adopts binary variables to represent students and exercises.  Guess and slip parameters are also introduced. 
\item \textbf{NCDM}~\cite{wang2020neural}. It applies neural networks to CD, and uses high-dimensional representations to represent students' abilities. 
\item \textbf{KaNCD}~\cite{wang2022neuralcd}.  Compared to NCDM, it adopts matrix factorization techniques for student abilities and exercise difficulties. 
\item \textbf{KSCD}~\cite{ma2022knowledge}. Compared to NCDM, it adopts matrix factorization techniques for student abilities and exercise difficulties, and designs a novel fusion function. 
\item \textbf{RCD}~\cite{gao2021rcd}. It incorporates student-exercise bipartite graph and concept-concept relationships into CD models. 
\item \textbf{SCD}~\cite{wang2023self}. Compared to RCD, it gives up diagnosing proficiency levels, and introduces a self-supervised learning based loss.   
\item \textbf{HAN-CD}~\cite{wang2019heterogeneous}. To adapt HAN to CD, we replace initial node features and projection with free embeddings, and consider correct/incorrect student-exercise meta paths.
\end{itemize}

For the sake of feasibility of KSCD with 24G cuda memory, we change its fusion function by directly fusing representations at the second dimension on Junyi and MOOC-Radar datasets. 
For RCD, we replace the GNN propagation process in the original codes with sparse tensor matrix multiplication, which  improves efficiency while keeping a similar diagnostic performance.  
Besides, we present a variation of edge differentiation layer in Appendix \ref{model_var}, and the corresponding results are recorded in Appendix \ref{sec:add_exp}.

\subsubsection{Hyperparameter Settings.} 
We randomly sample 8,192 logs per batch, and search the learning rate from \{0.0001, 0.0005, 0.001, 0.005, 0.01\} for all models. 
We adopt the Adam optimizer. We set the embedding dimension $Z$ to 128 for all models, and adopt the Xavier initialization for all trainable parameters. As for the number of GNN layers $L$, similar to previous studies~\cite{qian2024orcdf,wang2023self}, we search from \{1, 2, 3, 4\} and select the best result. 
The hidden dimensions of MLPs in Eq.(\ref{eq:diagnostic_function}) are  512 and 256 for all models. 
Following~\cite{yang2024graph}, the temperature {\small$t$} is set to 0.2, {\small$\alpha$} in Eq.(\ref{eq:RBF}) is set to 0.2 on all datasets. 
To ensure fair comparisons, for both ORCDF and our proposed \shortname, we choose KaNCD as the backbone model \cite{wang2022neuralcd}, and set the balancing hyperparameter $\beta$ to 0.5. Further, we adopt the same GNN structure for these two models.  
We have released codes of ISG-CD for implementation\footnote{https://github.com/ShaoPengyang/ISG-CD}. 

\begin{table}[ht]
    \centering
    \caption{Overall performance on the {ASSIST} dataset. We use bold font to emphasize the best results and underline to indicate the second-best results.} 
    \label{tab:assist}
    \scalebox{0.95}{
    \begin{tabular}{|c|c|c|c|}
    \hline
        \textbf{Model} & \textbf{ACC} $\uparrow$  & \textbf{AUC} $\uparrow$ & \textbf{DOA} $\uparrow$  \\ \hline
        \textbf{DINA} & 0.6253  $\pm$ 0.0245 & 0.6794  $\pm$ 0.0201 & 0.5579  $\pm$ 0.0316  \\ \hline
        \textbf{NCDM} & 0.7072  $\pm$ 0.0222 & 0.7244  $\pm$ 0.0212 & 0.5543 $\pm$ 0.0293  \\ \hline
        \textbf{KSCD} & 0.7209  $\pm$ 0.0241 & 0.7503  $\pm$ 0.0252 & 0.5092  $\pm$ 0.0062  \\ \hline
        \textbf{KaNCD} & 0.7182  $\pm$ 0.0250 & 0.7404  $\pm$ 0.0251 & 0.6057  $\pm$ 0.0235  \\ \hline
        \textbf{RCD} & 0.7153  $\pm$ 0.0123 & 0.7382  $\pm$ 0.0226 & 0.6221  $\pm$ 0.0230  \\ \hline
        \textbf{SCD} & 0.7212  $\pm$ 0.0566 & \underline{0.7552  $\pm$ 0.0576} & -  \\ \hline
        \textbf{HAN-CD} & \underline{0.7257  $\pm$ 0.0229} & 0.7524  $\pm$ 0.0240 & \underline{0.6348  $\pm$ 0.0185} \\ \hline
        \textbf{ISG-CD} & \textbf{0.7322  $\pm$ 0.0247} & \textbf{0.7604  $\pm$ 0.0296} & \textbf{0.6582  $\pm$ 0.0251} \\ \hline
    \end{tabular}}
\end{table}

\begin{table}[t]
    \centering
    \caption{Overall performance on the {Junyi} dataset. We use bold font to emphasize the best results and underline to indicate the second-best results.} 
    \label{tab:junyi}
    \scalebox{0.95}{
    \begin{tabular}{|c|c|c|c|}
    \hline
        \textbf{Model} & \textbf{ACC} $\uparrow$  & \textbf{AUC} $\uparrow$ & \textbf{DOA} $\uparrow$  \\ \hline
        \textbf{DINA} & 0.5105  $\pm$ 0.0230 & 0.6429  $\pm$ 0.0177 & 0.5002  $\pm$ 0.0056  \\ \hline
        \textbf{NCDM} & 0.7482  $\pm$ 0.0013 & 0.7816  $\pm$ 0.0013 & 0.5476 $\pm$ 0.0162  \\ \hline
        \textbf{KSCD} & 0.7561  $\pm$ 0.0029 & 0.7909  $\pm$ 0.0053 & 0.5001  $\pm$ 0.0039  \\ \hline
        \textbf{KaNCD} & 0.7536  $\pm$ 0.0020 & 0.7867  $\pm$ 0.0017 & 0.5529  $\pm$ 0.0212  \\ \hline
        \textbf{RCD} & 0.7512  $\pm$ 0.0044 & 0.7834  $\pm$ 0.0025 & 0.4996  $\pm$ 0.0050  \\ \hline
        \textbf{SCD} & 0.7576  $\pm$ 0.0056 & 0.7902  $\pm$ 0.0040 & -  \\ \hline
        \textbf{HAN-CD} & \underline{0.7626  $\pm$ 0.0039} & \underline{0.7957  $\pm$ 0.0080} & \underline{0.6469  $\pm$ 0.0132} \\ \hline
        \textbf{ISG-CD} & \textbf{0.7672  $\pm$ 0.0041} & \textbf{0.8058$\pm$ 0.0039} & \textbf{0.6728  $\pm$ 0.0208} \\ \hline
    \end{tabular}}
\end{table}

\begin{table}[t]
    \centering
    \caption{Overall performance on the {MOOC-Radar} dataset. We use bold font to emphasize the best results and underline to indicate the second-best results.} 
    \label{tab:mooc}
    \scalebox{0.95}{
    \begin{tabular}{|c|c|c|c|} 
    \hline
         \textbf{Model} & \textbf{ACC} $\uparrow$  & \textbf{AUC} $\uparrow$ & \textbf{DOA} $\uparrow$  \\ \hline
        \textbf{DINA} & 0.7792  $\pm$ 0.0047 & 0.8142  $\pm$ 0.0035 & 0.5436  $\pm$ 0.0141  \\ \hline
        \textbf{NCDM} & 0.8537  $\pm$ 0.0063 & 0.8663  $\pm$ 0.0056 & 0.6055 $\pm$ 0.0247  \\ \hline
        \textbf{KSCD} & 0.8587  $\pm$ 0.0060 & 0.8800  $\pm$ 0.0048 & 0.5027  $\pm$ 0.0139  \\ \hline
        \textbf{KaNCD} & 0.8576  $\pm$ 0.0093 & 0.8782  $\pm$ 0.0049 & 0.6980  $\pm$ 0.0161  \\ \hline
        \textbf{RCD} & 0.8599  $\pm$ 0.0062 & 0.8792  $\pm$ 0.0065 & 0.6678  $\pm$ 0.0121  \\ \hline
        \textbf{SCD} & \underline{0.8617  $\pm$ 0.0068} &  \underline{0.8810  $\pm$ 0.0051} & -  \\ \hline
        \textbf{HAN-CD} & 0.8600  $\pm$ 0.0037 & 0.8799  $\pm$ 0.0096 & \underline{0.7231  $\pm$ 0.0231} \\ \hline
        \textbf{ISG-CD} & \textbf{0.8676 $\pm$ 0.0045} & \textbf{0.8885  $\pm$ 0.0068} & \textbf{0.7432  $\pm$ 0.0132} \\ \hline
    \end{tabular}}
\end{table}

\subsection{Overall Performance (RQ1)}
\label{sec:overallperformance}
We report overall performance in Table \ref{tab:assist}, Table \ref{tab:junyi} and Table \ref{tab:mooc}. There are several observations from these three tables. 
\begin{itemize}[leftmargin=0.4cm]
    \item \textbf{First of all,  \shortname~has the best performance  on three datasets}. 
    On three datasets, \shortname~ achieves a stable improvement in accuracy performance compared to second-best results.  Furthermore, \shortname~has an improvement of nearly 1.5\% on the DOA metric on MOOC-Radar dataset. 
    This clearly demonstrates the stable and strong modeling ability of \shortname. 
    \item \textbf{Second, compared to graph-based CD models which do not distinguish edge heterogeneity (RCD, SCD), our proposed \shortname~achieves stable improvements on all three datasets.}
    Specifically, compared to RCD and SCD, \shortname~ has over 1\% accuracy improvements on ASSIST and Junyi datasets. 
    This prove that distinguishing edge semantics in CD will improve diagnosis performance of graph-based CD models. 
    \item \textbf{Third, we find that \shortname~ consistently outperforms both HAN-CD and ORCDF on three models}.  ompared to HAN-CD and ORCDF, \shortname~ has nearly 0.7\% AUC improvements on all three datasets. Note that,  HAN-CD, ORCDF and our proposed \shortname~ actually share a similar GNN structure. These results indicate the effectiveness of  constructing a reliable student-exercise graph structure in enhancing graph-based CD models. 
    \item \textbf{Last but not least, we find that graph based methods (HAN-CD, ORCDF, and \shortname) achieve the most accuracy improvements on the Junyi dataset compared to the backbone KaNCD model}. 
    The reason lies in the unique characteristic of Junyi dataset, i.e., one-to-one correspondence between knowledge concepts and exercises. 
    That is to say, concepts that we encountered in the testing set will be totally different from those in the training set. 
    This indicates that utilizing graph structure can better handle some extreme data situations. 
\end{itemize}

\subsection{Uncertain Edge Detection (RQ2)}
\label{sec:detect}
In this part, we focus on whether the IE-Diff layer can accurately distinguish uncertain edges. 
Note that, these uncertain edges arise due to students' carelessness or guessing. We find that it is not easy to generate random noise on non-interacted student-exercise pairs. These non-interacted pairs consist of potential correct and incorrect response logs. Hence, we randomly choose certain ratios of existing student-exercise response logs (corresponding to edges in graph) and change their labels. 
IE-Diff assigns {\small$Bern(w_{se}^0)$} or {\small$Bern(w_{se}^1)$} to the edge between student $s$ and exercise $e$. A smaller value of {\small$Bern(w_{se}^0)$} or {\small$Bern(w_{se}^1)$} denotes more uncertainty in this edge. If the value is smaller than 0.5, IE-Diff regards this edge as detected uncertain edges. We conduct experiments about whether these uncertain edges can be detected on ASSIST and Junyi datasets.

\begin{figure}[t]
    \begin{center}
    \includegraphics[width=0.5\textwidth]{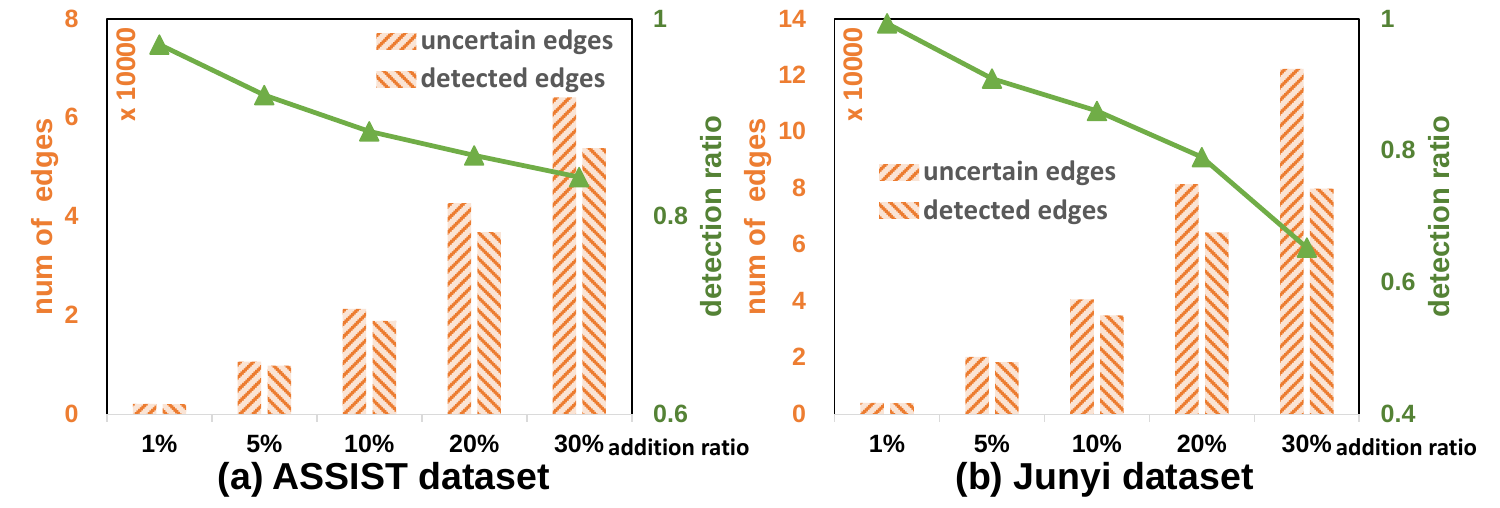}
    \end{center}
    \caption{Edge detection with varying the number of uncertain edges on the ASSIST and Junyi datasets. }
    \label{fig:detc}
\end{figure}

\begin{table*}[t]
    \centering
    \caption{Ablation studies on {ASSIST} and Junyi datasets. The last row corresponds to our proposed \shortname.} 
    \label{tab:ablation}
    \scalebox{0.85}{
    \begin{tabular}{|c|c|c|c|c|c|c|c|}
    \hline
        \textbf{~S-GNN~~} &
        \textbf{IE-Diff} &
        \multicolumn{3}{c|}{\textbf{ASSIST dataset}} & \multicolumn{3}{c|}{\textbf{Junyi dataset}}  \\ \cline{3-8}
        
         \textbf{Sec \ref{sec:model1}} &
         \textbf{Sec \ref{sec:model2}} & \textbf{ACC} $\uparrow$ &  \textbf{AUC} $\uparrow$ & \textbf{DOA} $\uparrow$ & \textbf{ACC} $\uparrow$ & \textbf{AUC} $\uparrow$ & \textbf{DOA} $\uparrow$ \\ \hline
        {\small\XSolidBrush} 
        &{\small\XSolidBrush}  
        & 0.7182  $\pm$ 0.0250 & 0.7404  $\pm$ 0.0251 & 0.6057  $\pm$ 0.0231  
        & 0.7536  $\pm$ 0.0020 & 0.7867  $\pm$ 0.0017 & 0.5529  $\pm$ 0.0212 \\ \hline
        \Checkmark
        &{\small\XSolidBrush} 
        & 0.7248  $\pm$ 0.0165 & 0.7518  $\pm$ 0.0188 & 0.6329  $\pm$ 0.0101  
        & 0.7623  $\pm$ 0.0082 & 0.7964  $\pm$ 0.0065 & 0.6432  $\pm$ 0.0123 \\ \hline
        \Checkmark 
        &\Checkmark
        &\textbf{0.7322  $\pm$ 0.0247} & \textbf{0.7604  $\pm$ 0.0296} & \textbf{0.6582  $\pm$ 0.0251}
        &\textbf{0.7672  $\pm$ 0.0041} & \textbf{0.8058$\pm$ 0.0041} & \textbf{0.6728  $\pm$ 0.0208} \\ \hline
    \end{tabular}}
\end{table*}

From Figure \ref{fig:detc}, we can infer two findings. First, the more modified edges, the lower the detection ratio. One possible reason is that the original data distribution will change if most edges are uncertain. Second, we find that even with a considerable number of uncertain edges, \shortname~still maintains relatively reliable detection performance. Specifically, on the ASSIST dataset, even with 30\% of the edges being uncertain, we still maintain a detection probability of near 80\%. These findings can reflect effectiveness of IE-Diff.

\subsection{Ablation Studies (RQ3)}
\label{sec:ablation}

We have conducted ablation experiments about the Semantic GNN (S-GNN) layer and Informative Edge Differentiation layer (IE-Diff) of \shortname, and results are presented in Table \ref{tab:ablation}. 
The first row does not take any additional layers into consideration, therefore, it corresponds to the backbone KaNCD model~\cite{wang2022neuralcd}. 
Please note that, S-GNN layer serves as the precondition for IE-Diff layer, therefore, we do not include adopting IE-Diff without S-GNN in  Table \ref{tab:ablation}.

There are several observations from Table \ref{tab:ablation}. 
First, Semantic-aware GNN Layer demonstrates remarkable adaptability to CD by combining state-of-the-art model paradigms~\cite{wang2020neural,wang2022neuralcd}. 
Compared to some models that sacrifice diagnostic capabilities for accuracy improvements~\cite{wang2023self}, it not only achieves a remarkable accuracy improvement but also retains its diagnostic capabilities intact. Second, Informative Edge Differentiation Layer makes precise adjustments to the graph structure, resulting in consistent accuracy improvements on both datasets. This layer builds on the aforementioned foundations, showcasing its ability to further enhance diagnostic performance.

\begin{figure}[t]
    \includegraphics[width=0.48\textwidth]{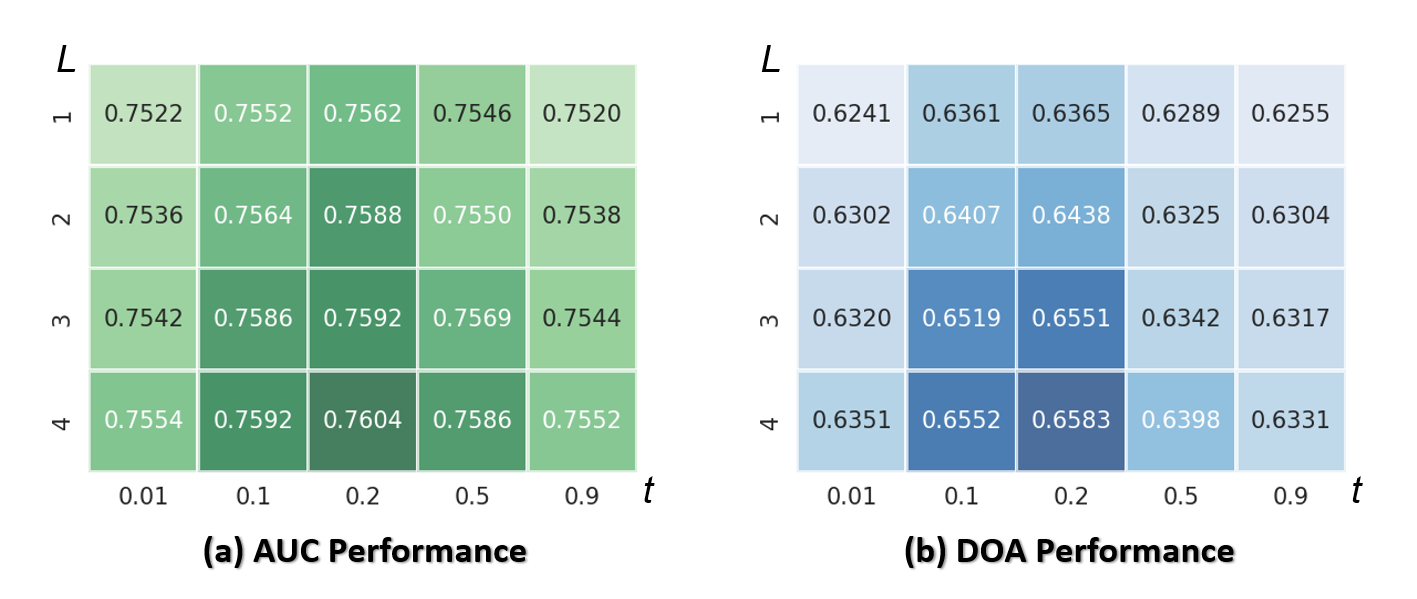}
    \caption{Diagnostic performance with varying number of GNN layers {$L$} and temperature $t$ on the ASSIST dataset. }
    \label{fig:layer}
\end{figure}

\subsection{Hyperparameters Analyses (RQ4)}
\label{sec:hyper}
\begin{figure}[t]
    \begin{center}
    \includegraphics[width=0.48\textwidth]{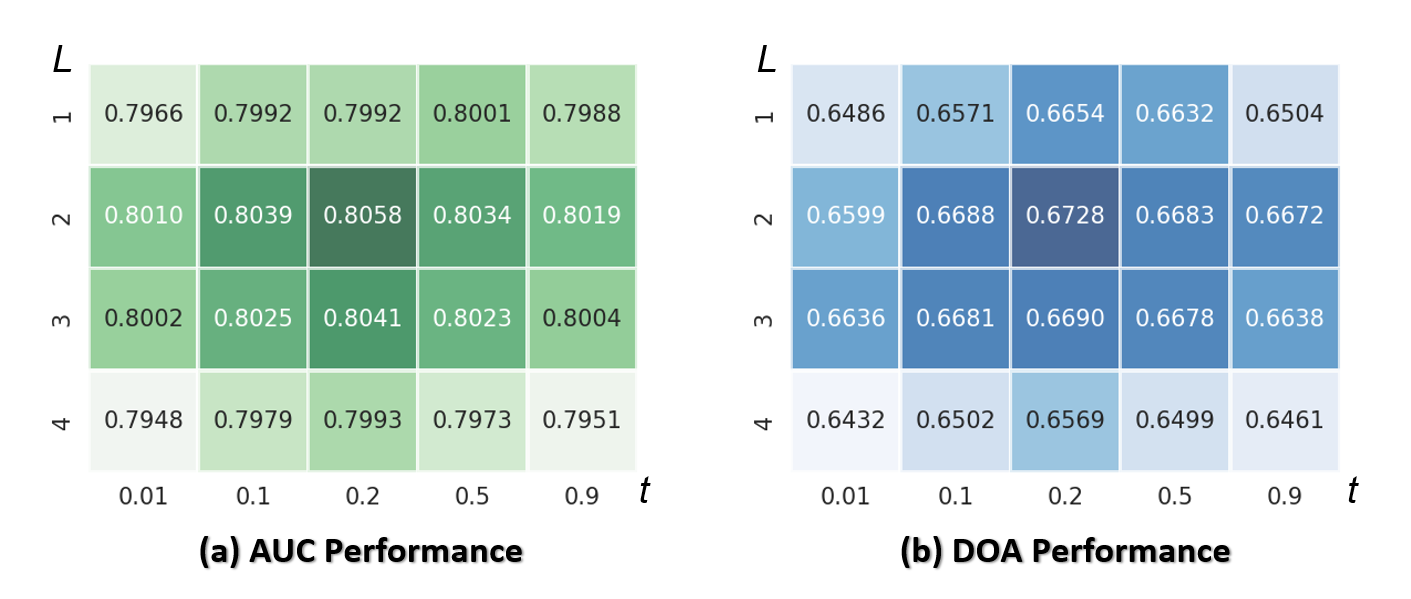}
    \end{center}
    \caption{Diagnostic performance with varying number of GNN layers {$L$} and temperature $t$ on the Junyi dataset. }
    \label{fig:layer-2}
\end{figure}

\textbf{Number of GNN Layers $L$ and Temperature $t$.} 
The number of GNN layers $L$ is directly related to performance of \shortname. 
Also, temperature $t$ in Eq.(\ref{eq:bern}) is an important hyperparameter that controls the effectiveness of $\mathcal{L}_{HSIC}$. In this part, we focus on studying the impacts of these two factors. The results are in Figure \ref{fig:layer}-\ref{fig:layer-2}. 

We have several observations from these results. 
First, we find that \shortname~ achieves a stable improvement compared to the backbone model, KaNCD. Specifically, AUC of KaNCD on ASSIST dataset is 0.7404. However, ACC of \shortname~ in Figure \ref{fig:layer} is over  0.7550 with ranging the number of GNN layers $L$ from 1 to 4. This indicates the stable effectiveness of our proposed semantic-aware GNN layers. 
Second, we find that the optimal results on the ASSIST dataset occur when $L=4$, while the optimal results on the Junyi dataset occur when $L=2$. The phenomenon on Junyi dataset indicates that there may be over-smoothing caused by multiple GNN layers.  These findings suggest that the number of GNN layers should be carefully chosen based on specific datasets. 
Third, when the temperature $t$ is in the range of 0.1 to 0.5, \shortname~ achieves relatively stable performance on both two datasets.  When the temperature $t$ approaches extreme values, \shortname~  experiences a noticeable decline in diagnosis performance. 
For example, on the ASSIST dataset,  AUC of \shortname~ is close to 0.76 when $t=0.1,0.2,0.5$ and $L=4$, while AUC is close to 0.755 when $t=0.01,0.9$ and $L=4$. These observation suggests that we should avoid extreme values when choosing temperature $t$.

\subsection{Comparing Different Training Strategies (RQ 5)}
\label{sec:strategies}

We compare the following training strategies: the Alternating Learning (AL)  strategy~\cite{zibetti2022alternating,zhu2024collaborative} and the Multi-Task Learning (MTL)  strategy~\cite{yang2024graph}. For the process of the AL strategy, we first pre-train {\small$\theta$} based on the original graph for 5 epochs. We also compare AL with ALwoP (AL without Pre-training). Subsequently, we alternately perform two steps: fix {\small$\theta$} to optimize {\small$\phi$} (Eq.(\ref{eq:final_loss})), then fix {\small$\phi$} to optimize {\small$\theta$} (Eq.(\ref{eq:BCE_loss})) until convergence. 
As for the process of the MTL strategy, we directly use Eq.(\ref{eq:final_loss}) to optimize overall parameters of \shortname~  together ({\small$\theta$} and {\small$\phi$})~\cite{yang2024graph}.

\begin{table}[t]
    \centering
    \caption{Comparing Different training strategies on the ASSIST and Junyi datasets. } 
    \label{tab:strategy}
    \scalebox{0.8}{
    \begin{tabular}{cc|ccc}
    \hline
        \textbf{Dataset} &\textbf{strategy} & \textbf{ACC $\uparrow$}  & \textbf{AUC $\uparrow$} & \textbf{DOA $\uparrow$}  \\ \hline
        \textbf{ASSIST} &\textbf{MTL} & {0.7271  $\pm$ 0.0186} & {0.7532  $\pm$ 0.0160} & {0.6299  $\pm$ 0.0211} \\ \hline
        \textbf{ASSIST} &\textbf{ALwoP} & {0.7301  $\pm$ 0.0284} & {0.7586  $\pm$ 0.0311} & {0.6402 $\pm$ 0.0268} \\ \hline
        \textbf{ASSIST} &\textbf{AL} & {0.7322  $\pm$ 0.0247} & {0.7604  $\pm$ 0.0296} & {0.6582  $\pm$ 0.0251} \\ \hline
        \textbf{Junyi} &\textbf{MTL} & {0.7621  $\pm$ 0.0066} & {0.8004 $\pm$ 0.0052} & {0.6524  $\pm$ 0.0137} \\ \hline
        \textbf{Junyi} &\textbf{ALwoP} & {0.7646  $\pm$ 0.0052} & {0.8028$\pm$ 0.0038} & {0.6633  $\pm$ 0.0162} \\ \hline
        \textbf{Junyi} &\textbf{AL} & {0.7672  $\pm$ 0.0041} & {0.8058$\pm$ 0.0039} & {0.6728  $\pm$ 0.0208} \\ \hline
    \end{tabular}}
\end{table}

The corresponding results are in Table \ref{tab:strategy}. Firstly,  the performance of MTL is the worst. A possible reason is that the parameters of the graph-based CD models ({\small$\theta$}) do not need to be constrained by HSIC loss, but MTL forces all parameters to be constrained by both BCE and HSIC. Secondly, we find that the pre-training stage can improve the performance of \shortname. We believe that the pre-training stage can enhance the stability of \shortname~ by providing better initial values of {\small$\theta$} for the following alternating training.

\section{Conclusion and Future Work}
Educational cognitive diagnosis is a critical task in the data mining domain, which focuses on diagnosing students' proficiency levels on knowledge concepts. 
In this paper, we proposed a novel \shortname, which captures the edge heterogeneity and uncertainty. 
Firstly, to capture edge heterogeneity, we designed the process of how to incorporate a semantic-aware GNN into CD. 
Secondly, we proposed an Informative Edge Differentiation layer, which focused on keeping a minimal yet sufficient reliable graph for CD based on the information bottleneck principle. 
Specifically, we formulated the process as goals of mutual information minimization/maximization, and converted these goals to different loss functions. Finally, we adopted alternating training to optimize the graph-based CD model and edge differentiation layer. 
Extensive experimental results on three datasets demonstrated the superiority of  \shortname. 
In the future, we plan to explore other solutions to  alleviating negative effects of uncertain edges.

\begin{acks}
This work has been supported in part by grants from the New Cornerstone Science Foundation through the XPLORER PRIZE, the National Science and Technology Major Project (2021ZD0111802), the National Natural Science Foundation of China (72188101, 62376086, U22A2094, 62272262). 
\end{acks}

\bibliographystyle{ACM-Reference-Format}
\bibliography{sample-base}


\begin{thebibliography}{59}


\ifx \showCODEN    \undefined \def \showCODEN     #1{\unskip}     \fi
\ifx \showISBNx    \undefined \def \showISBNx     #1{\unskip}     \fi
\ifx \showISBNxiii \undefined \def \showISBNxiii  #1{\unskip}     \fi
\ifx \showISSN     \undefined \def \showISSN      #1{\unskip}     \fi
\ifx \showLCCN     \undefined \def \showLCCN      #1{\unskip}     \fi
\ifx \shownote     \undefined \def \shownote      #1{#1}          \fi
\ifx \showarticletitle \undefined \def \showarticletitle #1{#1}   \fi
\ifx \showURL      \undefined \def \showURL       {\relax}        \fi
\providecommand\bibfield[2]{#2}
\providecommand\bibinfo[2]{#2}
\providecommand\natexlab[1]{#1}
\providecommand\showeprint[2][]{arXiv:#2}

\bibitem[Abdelaleem et~al\mbox{.}(2023)]%
        {abdelaleem2023deep}
\bibfield{author}{\bibinfo{person}{Eslam Abdelaleem}, \bibinfo{person}{Ilya Nemenman}, {and} \bibinfo{person}{K~Michael Martini}.} \bibinfo{year}{2023}\natexlab{}.
\newblock \showarticletitle{Deep Variational Multivariate Information Bottleneck--A Framework for Variational Losses}.
\newblock \bibinfo{journal}{\emph{arXiv preprint arXiv:2310.03311}} (\bibinfo{year}{2023}).
\newblock


\bibitem[Alemi et~al\mbox{.}(2016)]%
        {alemi2016deep}
\bibfield{author}{\bibinfo{person}{Alexander~A Alemi}, \bibinfo{person}{Ian Fischer}, \bibinfo{person}{Joshua~V Dillon}, {and} \bibinfo{person}{Kevin Murphy}.} \bibinfo{year}{2016}\natexlab{}.
\newblock \showarticletitle{Deep variational information bottleneck}.
\newblock \bibinfo{journal}{\emph{arXiv preprint arXiv:1612.00410}} (\bibinfo{year}{2016}).
\newblock


\bibitem[Bang et~al\mbox{.}(2021)]%
        {bang2021explaining}
\bibfield{author}{\bibinfo{person}{Seojin Bang}, \bibinfo{person}{Pengtao Xie}, \bibinfo{person}{Heewook Lee}, \bibinfo{person}{Wei Wu}, {and} \bibinfo{person}{Eric Xing}.} \bibinfo{year}{2021}\natexlab{}.
\newblock \showarticletitle{Explaining a black-box by using a deep variational information bottleneck approach}. In \bibinfo{booktitle}{\emph{Proceedings of the AAAI conference on artificial intelligence}}, Vol.~\bibinfo{volume}{35}. \bibinfo{pages}{11396--11404}.
\newblock


\bibitem[Berg et~al\mbox{.}(2017)]%
        {berg2017graph}
\bibfield{author}{\bibinfo{person}{Rianne van~den Berg}, \bibinfo{person}{Thomas~N Kipf}, {and} \bibinfo{person}{Max Welling}.} \bibinfo{year}{2017}\natexlab{}.
\newblock \showarticletitle{Graph convolutional matrix completion}.
\newblock \bibinfo{journal}{\emph{arXiv preprint arXiv:1706.02263}} (\bibinfo{year}{2017}).
\newblock


\bibitem[Berlinet and Thomas-Agnan(2011)]%
        {berlinet2011reproducing}
\bibfield{author}{\bibinfo{person}{Alain Berlinet} {and} \bibinfo{person}{Christine Thomas-Agnan}.} \bibinfo{year}{2011}\natexlab{}.
\newblock \bibinfo{booktitle}{\emph{Reproducing kernel Hilbert spaces in probability and statistics}}.
\newblock \bibinfo{publisher}{Springer Science \& Business Media}.
\newblock


\bibitem[Cai et~al\mbox{.}(2024)]%
        {cai2024popularity}
\bibfield{author}{\bibinfo{person}{Miaomiao Cai}, \bibinfo{person}{Lei Chen}, \bibinfo{person}{Yifan Wang}, \bibinfo{person}{Haoyue Bai}, \bibinfo{person}{Peijie Sun}, \bibinfo{person}{Le Wu}, \bibinfo{person}{Min Zhang}, {and} \bibinfo{person}{Meng Wang}.} \bibinfo{year}{2024}\natexlab{}.
\newblock \showarticletitle{Popularity-aware alignment and contrast for mitigating popularity bias}. In \bibinfo{booktitle}{\emph{Proceedings of the 30th ACM SIGKDD Conference on Knowledge Discovery and Data Mining}}. \bibinfo{pages}{187--198}.
\newblock


\bibitem[Chen et~al\mbox{.}(2023a)]%
        {chen2023adap}
\bibfield{author}{\bibinfo{person}{Jiawei Chen}, \bibinfo{person}{Junkang Wu}, \bibinfo{person}{Jiancan Wu}, \bibinfo{person}{Xuezhi Cao}, \bibinfo{person}{Sheng Zhou}, {and} \bibinfo{person}{Xiangnan He}.} \bibinfo{year}{2023}\natexlab{a}.
\newblock \showarticletitle{Adap-$\tau$: Adaptively Modulating Embedding Magnitude for Recommendation}. In \bibinfo{booktitle}{\emph{Proceedings of the ACM Web Conference 2023}}. \bibinfo{pages}{1085--1096}.
\newblock


\bibitem[Chen et~al\mbox{.}(2023b)]%
        {chen2023improving}
\bibfield{author}{\bibinfo{person}{Lei Chen}, \bibinfo{person}{Le Wu}, \bibinfo{person}{Kun Zhang}, \bibinfo{person}{Richang Hong}, \bibinfo{person}{Defu Lian}, \bibinfo{person}{Zhiqiang Zhang}, \bibinfo{person}{Jun Zhou}, {and} \bibinfo{person}{Meng Wang}.} \bibinfo{year}{2023}\natexlab{b}.
\newblock \showarticletitle{Improving recommendation fairness via data augmentation}. In \bibinfo{booktitle}{\emph{Proceedings of the ACM Web Conference 2023}}. \bibinfo{pages}{1012--1020}.
\newblock


\bibitem[Cui et~al\mbox{.}(2024)]%
        {cui2024leveraging}
\bibfield{author}{\bibinfo{person}{Jiajun Cui}, \bibinfo{person}{Hong Qian}, \bibinfo{person}{Bo Jiang}, {and} \bibinfo{person}{Wei Zhang}.} \bibinfo{year}{2024}\natexlab{}.
\newblock \showarticletitle{Leveraging Pedagogical Theories to Understand Student Learning Process with Graph-based Reasonable Knowledge Tracing}.
\newblock \bibinfo{journal}{\emph{arXiv preprint arXiv:2406.12896}} (\bibinfo{year}{2024}).
\newblock


\bibitem[De~La~Torre(2009)]%
        {de2009dina}
\bibfield{author}{\bibinfo{person}{Jimmy De~La~Torre}.} \bibinfo{year}{2009}\natexlab{}.
\newblock \showarticletitle{DINA model and parameter estimation: A didactic}.
\newblock \bibinfo{journal}{\emph{Journal of educational and behavioral statistics}} \bibinfo{volume}{34}, \bibinfo{number}{1} (\bibinfo{year}{2009}), \bibinfo{pages}{115--130}.
\newblock


\bibitem[Embretson and Reise(2013)]%
        {embretson2013item}
\bibfield{author}{\bibinfo{person}{Susan~E Embretson} {and} \bibinfo{person}{Steven~P Reise}.} \bibinfo{year}{2013}\natexlab{}.
\newblock \bibinfo{booktitle}{\emph{Item response theory}}.
\newblock \bibinfo{publisher}{Psychology Press}.
\newblock


\bibitem[Feng et~al\mbox{.}(2009)]%
        {feng2009addressing}
\bibfield{author}{\bibinfo{person}{Mingyu Feng}, \bibinfo{person}{Neil Heffernan}, {and} \bibinfo{person}{Kenneth Koedinger}.} \bibinfo{year}{2009}\natexlab{}.
\newblock \showarticletitle{Addressing the assessment challenge with an online system that tutors as it assesses}.
\newblock \bibinfo{journal}{\emph{User modeling and user-adapted interaction}}  \bibinfo{volume}{19} (\bibinfo{year}{2009}), \bibinfo{pages}{243--266}.
\newblock


\bibitem[Gao et~al\mbox{.}(2023a)]%
        {gao2023cirs}
\bibfield{author}{\bibinfo{person}{Chongming Gao}, \bibinfo{person}{Shiqi Wang}, \bibinfo{person}{Shijun Li}, \bibinfo{person}{Jiawei Chen}, \bibinfo{person}{Xiangnan He}, \bibinfo{person}{Wenqiang Lei}, \bibinfo{person}{Biao Li}, \bibinfo{person}{Yuan Zhang}, {and} \bibinfo{person}{Peng Jiang}.} \bibinfo{year}{2023}\natexlab{a}.
\newblock \showarticletitle{CIRS: Bursting filter bubbles by counterfactual interactive recommender system}.
\newblock \bibinfo{journal}{\emph{ACM Transactions on Information Systems}} \bibinfo{volume}{42}, \bibinfo{number}{1} (\bibinfo{year}{2023}), \bibinfo{pages}{1--27}.
\newblock


\bibitem[Gao et~al\mbox{.}(2021)]%
        {gao2021rcd}
\bibfield{author}{\bibinfo{person}{Weibo Gao}, \bibinfo{person}{Qi Liu}, \bibinfo{person}{Zhenya Huang}, \bibinfo{person}{Yu Yin}, \bibinfo{person}{Haoyang Bi}, \bibinfo{person}{Mu-Chun Wang}, \bibinfo{person}{Jianhui Ma}, \bibinfo{person}{Shijin Wang}, {and} \bibinfo{person}{Yu Su}.} \bibinfo{year}{2021}\natexlab{}.
\newblock \showarticletitle{RCD: Relation map driven cognitive diagnosis for intelligent education systems}. In \bibinfo{booktitle}{\emph{Proceedings of the 44th international ACM SIGIR conference on research and development in information retrieval}}. \bibinfo{pages}{501--510}.
\newblock


\bibitem[Gao et~al\mbox{.}(2023b)]%
        {gao2023leveraging}
\bibfield{author}{\bibinfo{person}{Weibo Gao}, \bibinfo{person}{Hao Wang}, \bibinfo{person}{Qi Liu}, \bibinfo{person}{Fei Wang}, \bibinfo{person}{Xin Lin}, \bibinfo{person}{Linan Yue}, \bibinfo{person}{Zheng Zhang}, \bibinfo{person}{Rui Lv}, {and} \bibinfo{person}{Shijin Wang}.} \bibinfo{year}{2023}\natexlab{b}.
\newblock \showarticletitle{Leveraging Transferable Knowledge Concept Graph Embedding for Cold-Start Cognitive Diagnosis}. In \bibinfo{booktitle}{\emph{Proceedings of the 46th International ACM SIGIR Conference on Research and Development in Information Retrieval}}. \bibinfo{pages}{983--992}.
\newblock


\bibitem[Gretton et~al\mbox{.}(2005)]%
        {gretton2005measuring}
\bibfield{author}{\bibinfo{person}{Arthur Gretton}, \bibinfo{person}{Olivier Bousquet}, \bibinfo{person}{Alex Smola}, {and} \bibinfo{person}{Bernhard Sch{\"o}lkopf}.} \bibinfo{year}{2005}\natexlab{}.
\newblock \showarticletitle{Measuring statistical dependence with Hilbert-Schmidt norms}. In \bibinfo{booktitle}{\emph{International conference on algorithmic learning theory}}. Springer, \bibinfo{pages}{63--77}.
\newblock


\bibitem[Hu et~al\mbox{.}(2020)]%
        {hu2020heterogeneous}
\bibfield{author}{\bibinfo{person}{Ziniu Hu}, \bibinfo{person}{Yuxiao Dong}, \bibinfo{person}{Kuansan Wang}, {and} \bibinfo{person}{Yizhou Sun}.} \bibinfo{year}{2020}\natexlab{}.
\newblock \showarticletitle{Heterogeneous graph transformer}. In \bibinfo{booktitle}{\emph{Proceedings of the web conference 2020}}. \bibinfo{pages}{2704--2710}.
\newblock


\bibitem[Jang et~al\mbox{.}(2016)]%
        {jang2016categorical}
\bibfield{author}{\bibinfo{person}{Eric Jang}, \bibinfo{person}{Shixiang Gu}, {and} \bibinfo{person}{Ben Poole}.} \bibinfo{year}{2016}\natexlab{}.
\newblock \showarticletitle{Categorical reparameterization with gumbel-softmax}.
\newblock \bibinfo{journal}{\emph{arXiv preprint arXiv:1611.01144}} (\bibinfo{year}{2016}).
\newblock


\bibitem[Li et~al\mbox{.}(2022)]%
        {li2022hiercdf}
\bibfield{author}{\bibinfo{person}{Jiatong Li}, \bibinfo{person}{Fei Wang}, \bibinfo{person}{Qi Liu}, \bibinfo{person}{Mengxiao Zhu}, \bibinfo{person}{Wei Huang}, \bibinfo{person}{Zhenya Huang}, \bibinfo{person}{Enhong Chen}, \bibinfo{person}{Yu Su}, {and} \bibinfo{person}{Shijin Wang}.} \bibinfo{year}{2022}\natexlab{}.
\newblock \showarticletitle{HierCDF: A Bayesian Network-based Hierarchical Cognitive Diagnosis Framework}. In \bibinfo{booktitle}{\emph{Proceedings of the 28th ACM SIGKDD Conference on Knowledge Discovery and Data Mining}}. \bibinfo{pages}{904--913}.
\newblock


\bibitem[Liu et~al\mbox{.}(2024)]%
        {liu2024inductive}
\bibfield{author}{\bibinfo{person}{Shuo Liu}, \bibinfo{person}{Junhao Shen}, \bibinfo{person}{Hong Qian}, {and} \bibinfo{person}{Aimin Zhou}.} \bibinfo{year}{2024}\natexlab{}.
\newblock \showarticletitle{Inductive Cognitive Diagnosis for Fast Student Learning in Web-Based Intelligent Education Systems}. In \bibinfo{booktitle}{\emph{Proceedings of the ACM on Web Conference 2024}}. \bibinfo{pages}{4260--4271}.
\newblock


\bibitem[Liu et~al\mbox{.}(2022)]%
        {liu2022agfa}
\bibfield{author}{\bibinfo{person}{Xinpu Liu}, \bibinfo{person}{Yanxin Ma}, \bibinfo{person}{Ke Xu}, \bibinfo{person}{Jianwei Wan}, {and} \bibinfo{person}{Yulan Guo}.} \bibinfo{year}{2022}\natexlab{}.
\newblock \showarticletitle{AGFA-Net: Adaptive global feature augmentation network for point cloud completion}.
\newblock \bibinfo{journal}{\emph{IEEE Geoscience and Remote Sensing Letters}}  \bibinfo{volume}{19} (\bibinfo{year}{2022}), \bibinfo{pages}{1--5}.
\newblock


\bibitem[Ma et~al\mbox{.}(2022)]%
        {ma2022knowledge}
\bibfield{author}{\bibinfo{person}{Haiping Ma}, \bibinfo{person}{Manwei Li}, \bibinfo{person}{Le Wu}, \bibinfo{person}{Haifeng Zhang}, \bibinfo{person}{Yunbo Cao}, \bibinfo{person}{Xingyi Zhang}, {and} \bibinfo{person}{Xuemin Zhao}.} \bibinfo{year}{2022}\natexlab{}.
\newblock \showarticletitle{Knowledge-Sensed Cognitive Diagnosis for Intelligent Education Platforms}. In \bibinfo{booktitle}{\emph{Proceedings of the 31st ACM International Conference on Information \& Knowledge Management}}. \bibinfo{pages}{1451--1460}.
\newblock


\bibitem[Ma et~al\mbox{.}(2024a)]%
        {ma2024dgcd}
\bibfield{author}{\bibinfo{person}{Haiping Ma}, \bibinfo{person}{Siyu Song}, \bibinfo{person}{Chuan Qin}, \bibinfo{person}{Xiaoshan Yu}, \bibinfo{person}{Limiao Zhang}, \bibinfo{person}{Xingyi Zhang}, {and} \bibinfo{person}{Hengshu Zhu}.} \bibinfo{year}{2024}\natexlab{a}.
\newblock \showarticletitle{DGCD: An Adaptive Denoising GNN for Group-level Cognitive Diagnosis}. In \bibinfo{booktitle}{\emph{The 33rd International Joint Conference on Artificial Intelligence (IJCAI-24)}}.
\newblock


\bibitem[Ma et~al\mbox{.}(2024b)]%
        {ma2024enhancing}
\bibfield{author}{\bibinfo{person}{Haiping Ma}, \bibinfo{person}{Changqian Wang}, \bibinfo{person}{Hengshu Zhu}, \bibinfo{person}{Shangshang Yang}, \bibinfo{person}{Xiaoming Zhang}, {and} \bibinfo{person}{Xingyi Zhang}.} \bibinfo{year}{2024}\natexlab{b}.
\newblock \showarticletitle{Enhancing cognitive diagnosis using un-interacted exercises: A collaboration-aware mixed sampling approach}. In \bibinfo{booktitle}{\emph{Proceedings of the AAAI Conference on Artificial Intelligence}}, Vol.~\bibinfo{volume}{38}. \bibinfo{pages}{8877--8885}.
\newblock


\bibitem[Ma et~al\mbox{.}(2020)]%
        {ma2020hsic}
\bibfield{author}{\bibinfo{person}{Wan-Duo~Kurt Ma}, \bibinfo{person}{JP Lewis}, {and} \bibinfo{person}{W~Bastiaan Kleijn}.} \bibinfo{year}{2020}\natexlab{}.
\newblock \showarticletitle{The HSIC bottleneck: Deep learning without back-propagation}. In \bibinfo{booktitle}{\emph{Proceedings of the AAAI conference on artificial intelligence}}, Vol.~\bibinfo{volume}{34}. \bibinfo{pages}{5085--5092}.
\newblock


\bibitem[Mnih and Salakhutdinov(2007)]%
        {mnih2007probabilistic}
\bibfield{author}{\bibinfo{person}{Andriy Mnih} {and} \bibinfo{person}{Russ~R Salakhutdinov}.} \bibinfo{year}{2007}\natexlab{}.
\newblock \showarticletitle{Probabilistic matrix factorization}.
\newblock \bibinfo{journal}{\emph{Advances in neural information processing systems}}  \bibinfo{volume}{20} (\bibinfo{year}{2007}).
\newblock


\bibitem[Qian et~al\mbox{.}(2024)]%
        {qian2024orcdf}
\bibfield{author}{\bibinfo{person}{Hong Qian}, \bibinfo{person}{Shuo Liu}, \bibinfo{person}{Mingjia Li}, \bibinfo{person}{Bingdong Li}, \bibinfo{person}{Zhi Liu}, {and} \bibinfo{person}{Aimin Zhou}.} \bibinfo{year}{2024}\natexlab{}.
\newblock \showarticletitle{ORCDF: An Oversmoothing-Resistant Cognitive Diagnosis Framework for Student Learning in Online Education Systems}. In \bibinfo{booktitle}{\emph{Proceedings of the 30th ACM SIGKDD Conference on Knowledge Discovery and Data Mining}}. \bibinfo{pages}{2455--2466}.
\newblock


\bibitem[Reckase(1997)]%
        {reckase1997past}
\bibfield{author}{\bibinfo{person}{Mark~D Reckase}.} \bibinfo{year}{1997}\natexlab{}.
\newblock \showarticletitle{The past and future of multidimensional item response theory}.
\newblock \bibinfo{journal}{\emph{Applied Psychological Measurement}} \bibinfo{volume}{21}, \bibinfo{number}{1} (\bibinfo{year}{1997}), \bibinfo{pages}{25--36}.
\newblock


\bibitem[Saxe et~al\mbox{.}(2019)]%
        {saxe2019information}
\bibfield{author}{\bibinfo{person}{Andrew~M Saxe}, \bibinfo{person}{Yamini Bansal}, \bibinfo{person}{Joel Dapello}, \bibinfo{person}{Madhu Advani}, \bibinfo{person}{Artemy Kolchinsky}, \bibinfo{person}{Brendan~D Tracey}, {and} \bibinfo{person}{David~D Cox}.} \bibinfo{year}{2019}\natexlab{}.
\newblock \showarticletitle{On the information bottleneck theory of deep learning}.
\newblock \bibinfo{journal}{\emph{Journal of Statistical Mechanics: Theory and Experiment}} \bibinfo{volume}{2019}, \bibinfo{number}{12} (\bibinfo{year}{2019}), \bibinfo{pages}{124020}.
\newblock


\bibitem[Shao et~al\mbox{.}({[n.\,d.]})]%
        {FCS2024shao}
\bibfield{author}{\bibinfo{person}{Pengyang Shao}, \bibinfo{person}{Kun Zhang}, \bibinfo{person}{Chen Gao}, \bibinfo{person}{Lei Chen}, \bibinfo{person}{Miaomiao Cai}, \bibinfo{person}{Le Wu}, \bibinfo{person}{Yong Li}, {and} \bibinfo{person}{Meng Wang}.} \bibinfo{year}{[n.\,d.]}\natexlab{}.
\newblock \showarticletitle{Breaking Student-Concept Sparsity Barrier for Cognitive Diagnosis}.
\newblock \bibinfo{journal}{\emph{Frontiers of Computer Science}} (\bibinfo{year}{[n.\,d.]}).
\newblock
\href{https://doi.org/10.1007/s11704-025-40591-2}{doi:\nolinkurl{10.1007/s11704-025-40591-2}}


\bibitem[Shen et~al\mbox{.}(2024)]%
        {shen2024capturing}
\bibfield{author}{\bibinfo{person}{Junhao Shen}, \bibinfo{person}{Hong Qian}, \bibinfo{person}{Shuo Liu}, \bibinfo{person}{Wei Zhang}, \bibinfo{person}{Bo Jiang}, {and} \bibinfo{person}{Aimin Zhou}.} \bibinfo{year}{2024}\natexlab{}.
\newblock \showarticletitle{Capturing Homogeneous Influence among Students: Hypergraph Cognitive Diagnosis for Intelligent Education Systems}. In \bibinfo{booktitle}{\emph{Proceedings of the 30th ACM SIGKDD Conference on Knowledge Discovery and Data Mining}}. \bibinfo{pages}{2628--2639}.
\newblock


\bibitem[Sun et~al\mbox{.}(2022a)]%
        {sun2022adversarial}
\bibfield{author}{\bibinfo{person}{Lichao Sun}, \bibinfo{person}{Yingtong Dou}, \bibinfo{person}{Carl Yang}, \bibinfo{person}{Kai Zhang}, \bibinfo{person}{Ji Wang}, \bibinfo{person}{S~Yu Philip}, \bibinfo{person}{Lifang He}, {and} \bibinfo{person}{Bo Li}.} \bibinfo{year}{2022}\natexlab{a}.
\newblock \showarticletitle{Adversarial attack and defense on graph data: A survey}.
\newblock \bibinfo{journal}{\emph{IEEE Transactions on Knowledge and Data Engineering}} \bibinfo{volume}{35}, \bibinfo{number}{8} (\bibinfo{year}{2022}), \bibinfo{pages}{7693--7711}.
\newblock


\bibitem[Sun et~al\mbox{.}(2022b)]%
        {sun2022graph}
\bibfield{author}{\bibinfo{person}{Qingyun Sun}, \bibinfo{person}{Jianxin Li}, \bibinfo{person}{Hao Peng}, \bibinfo{person}{Jia Wu}, \bibinfo{person}{Xingcheng Fu}, \bibinfo{person}{Cheng Ji}, {and} \bibinfo{person}{S~Yu Philip}.} \bibinfo{year}{2022}\natexlab{b}.
\newblock \showarticletitle{Graph structure learning with variational information bottleneck}. In \bibinfo{booktitle}{\emph{Proceedings of the AAAI Conference on Artificial Intelligence}}, Vol.~\bibinfo{volume}{36}. \bibinfo{pages}{4165--4174}.
\newblock


\bibitem[Tishby and Zaslavsky(2015)]%
        {tishby2015deep}
\bibfield{author}{\bibinfo{person}{Naftali Tishby} {and} \bibinfo{person}{Noga Zaslavsky}.} \bibinfo{year}{2015}\natexlab{}.
\newblock \showarticletitle{Deep learning and the information bottleneck principle}. In \bibinfo{booktitle}{\emph{2015 ieee information theory workshop (itw)}}. IEEE, \bibinfo{pages}{1--5}.
\newblock


\bibitem[Vert et~al\mbox{.}(2004)]%
        {vert2004primer}
\bibfield{author}{\bibinfo{person}{Jean-Philippe Vert}, \bibinfo{person}{Koji Tsuda}, {and} \bibinfo{person}{Bernhard Sch{\"o}lkopf}.} \bibinfo{year}{2004}\natexlab{}.
\newblock \showarticletitle{A primer on kernel methods}.
\newblock  (\bibinfo{year}{2004}).
\newblock


\bibitem[Wang et~al\mbox{.}(2024a)]%
        {wang2024survey}
\bibfield{author}{\bibinfo{person}{Fei Wang}, \bibinfo{person}{Weibo Gao}, \bibinfo{person}{Qi Liu}, \bibinfo{person}{Jiatong Li}, \bibinfo{person}{Guanhao Zhao}, \bibinfo{person}{Zheng Zhang}, \bibinfo{person}{Zhenya Huang}, \bibinfo{person}{Mengxiao Zhu}, \bibinfo{person}{Shijin Wang}, \bibinfo{person}{Wei Tong}, {et~al\mbox{.}}} \bibinfo{year}{2024}\natexlab{a}.
\newblock \showarticletitle{A survey of models for cognitive diagnosis: New developments and future directions}.
\newblock \bibinfo{journal}{\emph{arXiv preprint arXiv:2407.05458}} (\bibinfo{year}{2024}).
\newblock


\bibitem[Wang et~al\mbox{.}(2023a)]%
        {wang2023dynamic}
\bibfield{author}{\bibinfo{person}{Fei Wang}, \bibinfo{person}{Zhenya Huang}, \bibinfo{person}{Qi Liu}, \bibinfo{person}{Enhong Chen}, \bibinfo{person}{Yu Yin}, \bibinfo{person}{Jianhui Ma}, {and} \bibinfo{person}{Shijin Wang}.} \bibinfo{year}{2023}\natexlab{a}.
\newblock \showarticletitle{Dynamic Cognitive Diagnosis: An Educational Priors-Enhanced Deep Knowledge Tracing Perspective}.
\newblock \bibinfo{journal}{\emph{IEEE Transactions on Learning Technologies}} (\bibinfo{year}{2023}).
\newblock


\bibitem[Wang et~al\mbox{.}(2020)]%
        {wang2020neural}
\bibfield{author}{\bibinfo{person}{Fei Wang}, \bibinfo{person}{Qi Liu}, \bibinfo{person}{Enhong Chen}, \bibinfo{person}{Zhenya Huang}, \bibinfo{person}{Yuying Chen}, \bibinfo{person}{Yu Yin}, \bibinfo{person}{Zai Huang}, {and} \bibinfo{person}{Shijin Wang}.} \bibinfo{year}{2020}\natexlab{}.
\newblock \showarticletitle{Neural cognitive diagnosis for intelligent education systems}. In \bibinfo{booktitle}{\emph{Proceedings of the AAAI conference on artificial intelligence}}, Vol.~\bibinfo{volume}{34}. \bibinfo{pages}{6153--6161}.
\newblock


\bibitem[Wang et~al\mbox{.}(2022)]%
        {wang2022neuralcd}
\bibfield{author}{\bibinfo{person}{Fei Wang}, \bibinfo{person}{Qi Liu}, \bibinfo{person}{Enhong Chen}, \bibinfo{person}{Zhenya Huang}, \bibinfo{person}{Yu Yin}, \bibinfo{person}{Shijin Wang}, {and} \bibinfo{person}{Yu Su}.} \bibinfo{year}{2022}\natexlab{}.
\newblock \showarticletitle{NeuralCD: a general framework for cognitive diagnosis}.
\newblock \bibinfo{journal}{\emph{IEEE Transactions on Knowledge and Data Engineering}} (\bibinfo{year}{2022}).
\newblock


\bibitem[Wang et~al\mbox{.}(2023b)]%
        {wang2023toward}
\bibfield{author}{\bibinfo{person}{Jihong Wang}, \bibinfo{person}{Minnan Luo}, \bibinfo{person}{Jundong Li}, \bibinfo{person}{Ziqi Liu}, \bibinfo{person}{Jun Zhou}, {and} \bibinfo{person}{Qinghua Zheng}.} \bibinfo{year}{2023}\natexlab{b}.
\newblock \showarticletitle{Toward enhanced robustness in unsupervised graph representation learning: A graph information bottleneck perspective}.
\newblock \bibinfo{journal}{\emph{IEEE Transactions on Knowledge and Data Engineering}} (\bibinfo{year}{2023}).
\newblock


\bibitem[Wang et~al\mbox{.}(2024b)]%
        {wang2024boosting}
\bibfield{author}{\bibinfo{person}{Shanshan Wang}, \bibinfo{person}{Zhen Zeng}, \bibinfo{person}{Xun Yang}, \bibinfo{person}{Ke Xu}, {and} \bibinfo{person}{Xingyi Zhang}.} \bibinfo{year}{2024}\natexlab{b}.
\newblock \showarticletitle{Boosting neural cognitive diagnosis with student’s affective state modeling}. In \bibinfo{booktitle}{\emph{Proceedings of the AAAI Conference on Artificial Intelligence}}, Vol.~\bibinfo{volume}{38}. \bibinfo{pages}{620--627}.
\newblock


\bibitem[Wang et~al\mbox{.}(2023c)]%
        {wang2023self}
\bibfield{author}{\bibinfo{person}{Shanshan Wang}, \bibinfo{person}{Zhen Zeng}, \bibinfo{person}{Xun Yang}, {and} \bibinfo{person}{Xingyi Zhang}.} \bibinfo{year}{2023}\natexlab{c}.
\newblock \showarticletitle{Self-supervised Graph Learning for Long-tailed Cognitive Diagnosis}. In \bibinfo{booktitle}{\emph{Proceedings of the AAAI Conference on Artificial Intelligence}}, Vol.~\bibinfo{volume}{37}. \bibinfo{pages}{110--118}.
\newblock


\bibitem[Wang et~al\mbox{.}(2021a)]%
        {wang2021learning}
\bibfield{author}{\bibinfo{person}{Tinghua Wang}, \bibinfo{person}{Xiaolu Dai}, {and} \bibinfo{person}{Yuze Liu}.} \bibinfo{year}{2021}\natexlab{a}.
\newblock \showarticletitle{Learning with Hilbert--Schmidt independence criterion: A review and new perspectives}.
\newblock \bibinfo{journal}{\emph{Knowledge-based systems}}  \bibinfo{volume}{234} (\bibinfo{year}{2021}), \bibinfo{pages}{107567}.
\newblock


\bibitem[Wang et~al\mbox{.}(2019)]%
        {wang2019heterogeneous}
\bibfield{author}{\bibinfo{person}{Xiao Wang}, \bibinfo{person}{Houye Ji}, \bibinfo{person}{Chuan Shi}, \bibinfo{person}{Bai Wang}, \bibinfo{person}{Yanfang Ye}, \bibinfo{person}{Peng Cui}, {and} \bibinfo{person}{Philip~S Yu}.} \bibinfo{year}{2019}\natexlab{}.
\newblock \showarticletitle{Heterogeneous graph attention network}. In \bibinfo{booktitle}{\emph{The world wide web conference}}. \bibinfo{pages}{2022--2032}.
\newblock


\bibitem[Wang et~al\mbox{.}(2021b)]%
        {wang2021revisiting}
\bibfield{author}{\bibinfo{person}{Zifeng Wang}, \bibinfo{person}{Tong Jian}, \bibinfo{person}{Aria Masoomi}, \bibinfo{person}{Stratis Ioannidis}, {and} \bibinfo{person}{Jennifer Dy}.} \bibinfo{year}{2021}\natexlab{b}.
\newblock \showarticletitle{Revisiting hilbert-schmidt information bottleneck for adversarial robustness}.
\newblock \bibinfo{journal}{\emph{Advances in Neural Information Processing Systems}}  \bibinfo{volume}{34} (\bibinfo{year}{2021}), \bibinfo{pages}{586--597}.
\newblock


\bibitem[Wu et~al\mbox{.}(2023)]%
        {wu2023causality}
\bibfield{author}{\bibinfo{person}{Chenwang Wu}, \bibinfo{person}{Xiting Wang}, \bibinfo{person}{Defu Lian}, \bibinfo{person}{Xing Xie}, {and} \bibinfo{person}{Enhong Chen}.} \bibinfo{year}{2023}\natexlab{}.
\newblock \showarticletitle{A causality inspired framework for model interpretation}. In \bibinfo{booktitle}{\emph{Proceedings of the 29th ACM SIGKDD Conference on Knowledge Discovery and Data Mining}}. \bibinfo{pages}{2731--2741}.
\newblock


\bibitem[Wu et~al\mbox{.}(2024)]%
        {wu19edustudio}
\bibfield{author}{\bibinfo{person}{Le Wu}, \bibinfo{person}{Xiangzhi Chen}, \bibinfo{person}{Fei Liu}, \bibinfo{person}{Junsong Xie}, \bibinfo{person}{Chenao Xia}, \bibinfo{person}{Zhengtao Tan}, \bibinfo{person}{Mi Tian}, \bibinfo{person}{Jinglong Li}, \bibinfo{person}{Kun Zhang}, \bibinfo{person}{Defu Lian}, \bibinfo{person}{Hong Richang}, {and} \bibinfo{person}{Wang Meng}.} \bibinfo{year}{2024}\natexlab{}.
\newblock \showarticletitle{EduStudio: towards a unified library for student cognitive modeling}.
\newblock \bibinfo{journal}{\emph{Frontiers of Computer Science}} \bibinfo{volume}{19}, \bibinfo{number}{8} (\bibinfo{year}{2024}), \bibinfo{pages}{198342}.
\newblock


\bibitem[Wu et~al\mbox{.}(2020b)]%
        {wu2020joint}
\bibfield{author}{\bibinfo{person}{Le Wu}, \bibinfo{person}{Yonghui Yang}, \bibinfo{person}{Kun Zhang}, \bibinfo{person}{Richang Hong}, \bibinfo{person}{Yanjie Fu}, {and} \bibinfo{person}{Meng Wang}.} \bibinfo{year}{2020}\natexlab{b}.
\newblock \showarticletitle{Joint item recommendation and attribute inference: An adaptive graph convolutional network approach}. In \bibinfo{booktitle}{\emph{Proceedings of the 43rd International ACM SIGIR conference on research and development in Information Retrieval}}. \bibinfo{pages}{679--688}.
\newblock


\bibitem[Wu et~al\mbox{.}(2020a)]%
        {wu2020graph}
\bibfield{author}{\bibinfo{person}{Tailin Wu}, \bibinfo{person}{Hongyu Ren}, \bibinfo{person}{Pan Li}, {and} \bibinfo{person}{Jure Leskovec}.} \bibinfo{year}{2020}\natexlab{a}.
\newblock \showarticletitle{Graph information bottleneck}.
\newblock \bibinfo{journal}{\emph{Advances in Neural Information Processing Systems}}  \bibinfo{volume}{33} (\bibinfo{year}{2020}), \bibinfo{pages}{20437--20448}.
\newblock


\bibitem[Yang et~al\mbox{.}(2024a)]%
        {yang2024disengcd}
\bibfield{author}{\bibinfo{person}{Shangshang Yang}, \bibinfo{person}{Mingyang Chen}, \bibinfo{person}{Ziwen Wang}, \bibinfo{person}{Xiaoshan Yu}, \bibinfo{person}{Panpan Zhang}, \bibinfo{person}{Haiping Ma}, {and} \bibinfo{person}{Xingyi Zhang}.} \bibinfo{year}{2024}\natexlab{a}.
\newblock \showarticletitle{DisenGCD: A Meta Multigraph-assisted Disentangled Graph Learning Framework for Cognitive Diagnosis}.
\newblock \bibinfo{journal}{\emph{arXiv preprint arXiv:2410.17564}} (\bibinfo{year}{2024}).
\newblock


\bibitem[Yang et~al\mbox{.}(2021)]%
        {yang2021enhanced}
\bibfield{author}{\bibinfo{person}{Yonghui Yang}, \bibinfo{person}{Le Wu}, \bibinfo{person}{Richang Hong}, \bibinfo{person}{Kun Zhang}, {and} \bibinfo{person}{Meng Wang}.} \bibinfo{year}{2021}\natexlab{}.
\newblock \showarticletitle{Enhanced graph learning for collaborative filtering via mutual information maximization}. In \bibinfo{booktitle}{\emph{Proceedings of the 44th International ACM SIGIR Conference on Research and Development in Information Retrieval}}. \bibinfo{pages}{71--80}.
\newblock


\bibitem[Yang et~al\mbox{.}(2024b)]%
        {yang2024graph}
\bibfield{author}{\bibinfo{person}{Yonghui Yang}, \bibinfo{person}{Le Wu}, \bibinfo{person}{Zihan Wang}, \bibinfo{person}{Zhuangzhuang He}, \bibinfo{person}{Richang Hong}, {and} \bibinfo{person}{Meng Wang}.} \bibinfo{year}{2024}\natexlab{b}.
\newblock \showarticletitle{Graph Bottlenecked Social Recommendation}.
\newblock \bibinfo{journal}{\emph{arXiv preprint arXiv:2406.08214}} (\bibinfo{year}{2024}).
\newblock


\bibitem[Yu et~al\mbox{.}(2023)]%
        {yu2023moocradar}
\bibfield{author}{\bibinfo{person}{Jifan Yu}, \bibinfo{person}{Mengying Lu}, \bibinfo{person}{Qingyang Zhong}, \bibinfo{person}{Zijun Yao}, \bibinfo{person}{Shangqing Tu}, \bibinfo{person}{Zhengshan Liao}, \bibinfo{person}{Xiaoya Li}, \bibinfo{person}{Manli Li}, \bibinfo{person}{Lei Hou}, \bibinfo{person}{Hai-Tao Zheng}, {et~al\mbox{.}}} \bibinfo{year}{2023}\natexlab{}.
\newblock \showarticletitle{Moocradar: A fine-grained and multi-aspect knowledge repository for improving cognitive student modeling in moocs}. In \bibinfo{booktitle}{\emph{Proceedings of the 46th International ACM SIGIR Conference on Research and Development in Information Retrieval}}. \bibinfo{pages}{2924--2934}.
\newblock


\bibitem[Zhang et~al\mbox{.}(2024b)]%
        {zhang2024path}
\bibfield{author}{\bibinfo{person}{Dacao Zhang}, \bibinfo{person}{Kun Zhang}, \bibinfo{person}{Le Wu}, \bibinfo{person}{Mi Tian}, \bibinfo{person}{Richang Hong}, {and} \bibinfo{person}{Meng Wang}.} \bibinfo{year}{2024}\natexlab{b}.
\newblock \showarticletitle{Path-Specific Causal Reasoning for Fairness-aware Cognitive Diagnosis}. In \bibinfo{booktitle}{\emph{Proceedings of the 30th ACM SIGKDD Conference on Knowledge Discovery and Data Mining}}. \bibinfo{pages}{4143--4154}.
\newblock


\bibitem[Zhang et~al\mbox{.}(2022)]%
        {zhang2022graph}
\bibfield{author}{\bibinfo{person}{Yifei Zhang}, \bibinfo{person}{Hao Zhu}, \bibinfo{person}{Ziqiao Meng}, \bibinfo{person}{Piotr Koniusz}, {and} \bibinfo{person}{Irwin King}.} \bibinfo{year}{2022}\natexlab{}.
\newblock \showarticletitle{Graph-adaptive rectified linear unit for graph neural networks}. In \bibinfo{booktitle}{\emph{Proceedings of the ACM Web Conference 2022}}. \bibinfo{pages}{1331--1339}.
\newblock


\bibitem[Zhang et~al\mbox{.}(2023)]%
        {zhang2023fairlisa}
\bibfield{author}{\bibinfo{person}{Zheng Zhang}, \bibinfo{person}{Qi Liu}, \bibinfo{person}{Hao Jiang}, \bibinfo{person}{Fei Wang}, \bibinfo{person}{Yan Zhuang}, \bibinfo{person}{Le Wu}, \bibinfo{person}{Weibo Gao}, {and} \bibinfo{person}{Enhong Chen}.} \bibinfo{year}{2023}\natexlab{}.
\newblock \showarticletitle{Fairlisa: Fair user modeling with limited sensitive attributes information}. In \bibinfo{booktitle}{\emph{Thirty-seventh Conference on Neural Information Processing Systems}}.
\newblock


\bibitem[Zhang et~al\mbox{.}(2024a)]%
        {zhang2024understanding}
\bibfield{author}{\bibinfo{person}{Zheng Zhang}, \bibinfo{person}{Le Wu}, \bibinfo{person}{Qi Liu}, \bibinfo{person}{Jiayu Liu}, \bibinfo{person}{Zhenya Huang}, \bibinfo{person}{Yu Yin}, \bibinfo{person}{Yan Zhuang}, \bibinfo{person}{Weibo Gao}, {and} \bibinfo{person}{Enhong Chen}.} \bibinfo{year}{2024}\natexlab{a}.
\newblock \showarticletitle{Understanding and improving fairness in cognitive diagnosis}.
\newblock \bibinfo{journal}{\emph{Science China Information Sciences}} \bibinfo{volume}{67}, \bibinfo{number}{5} (\bibinfo{year}{2024}), \bibinfo{pages}{152106}.
\newblock


\bibitem[Zhu et~al\mbox{.}(2024)]%
        {zhu2024collaborative}
\bibfield{author}{\bibinfo{person}{Yaochen Zhu}, \bibinfo{person}{Liang Wu}, \bibinfo{person}{Qi Guo}, \bibinfo{person}{Liangjie Hong}, {and} \bibinfo{person}{Jundong Li}.} \bibinfo{year}{2024}\natexlab{}.
\newblock \showarticletitle{Collaborative large language model for recommender systems}. In \bibinfo{booktitle}{\emph{Proceedings of the ACM on Web Conference 2024}}. \bibinfo{pages}{3162--3172}.
\newblock


\bibitem[Zibetti et~al\mbox{.}(2022)]%
        {zibetti2022alternating}
\bibfield{author}{\bibinfo{person}{Marcelo Victor~Wust Zibetti}, \bibinfo{person}{Florian Knoll}, {and} \bibinfo{person}{Ravinder~R Regatte}.} \bibinfo{year}{2022}\natexlab{}.
\newblock \showarticletitle{Alternating learning approach for variational networks and undersampling pattern in parallel MRI applications}.
\newblock \bibinfo{journal}{\emph{IEEE transactions on computational imaging}}  \bibinfo{volume}{8} (\bibinfo{year}{2022}), \bibinfo{pages}{449--461}.
\newblock


\end{thebibliography}

\appendix

\section{More Details}

\subsection{Variation of Edge Differentiation: Ada-Diff}
\label{model_var}
Edge differentiation focuses on how to distinguish fortunate guesses (careless mistakes) from mastery (no mastery). This is crucial for GNNs because these uncertain edges may propagate incorrect information during aggregation. 
We have introduced IE-Diff based on the information bottleneck principle. Here, we introduce an intuitive approach, which is based on \emph{A}daptive Edge \emph{Diff}erentiation (\textbf{Ada-Diff}), formulated as: 
\begin{small}
    \begin{equation}
        \label{eq:adaptive}
    \begin{aligned}
     \tilde{\mathcal{G}}_1=\{(s,e)|{\hat{r}_{se}^{\tilde{\mathcal{G}}} > \Delta \land r_{se} =1}  \}, \tilde{\mathcal{G}}_0=\{(s,e)|{\hat{r}_{se}^{\tilde{\mathcal{G}}} < 1-\Delta \land r_{se} =0}  \}, 
    \end{aligned}
    \end{equation}
\end{small}
where {\small$\Delta$} denotes the threshold for adaptively distinguishing uncertain edges.  For a specific concept, if a student answers corresponding exercises correctly multiple times and only answers incorrectly once,  the incorrect answer will be considered as an uncertain edge. 

\subsection{Notation Table}
\label{notation}
For readability, we list important notations in Table \ref{tab:notations}.
\begin{table}[t]
    \centering
    \caption{Important Notations in \shortname. } 
    \label{tab:notations}
    \scalebox{0.78}{
    \begin{tabular}{c|c}
    \hline
    \multicolumn{2}{c}{Task Formulation} \\  \hline
    $S,E,K$ & student, exercise, concept entities \\
    $M,N,T$ & number of students, exercises, concepts \\ 
    $r_{se}$ & student$s$'s response log to exercise $e$ \\ 
    $R={(s,e,r_{se})}$ & triplets of students, exercises and response logs \\ 
    $\mathbf{Q}=\{q_{ek}\}^{N\times T}$ & relations between exercises and concepts \\
    $\mathbf{A}, \mathbf{a}_s$ &student abilities, student $s$'s ability \\
    $\mathbf{D}, \mathbf{d}_e$ &exercise difficulties, exercise $e$'s difficulty \\
    $a_{sk}$&student $s$'s ability on concept $k$ \\
    $d_{ek}$&exercise $e$'s difficulty on concept $k$  \\
    $\mathbf{h}^{disc}, h_e^{disc}$ &exercise discriminations, exercise $e$'s discrimination \\ \hline
    
    \multicolumn{2}{c}{Semantic-aware GNN (S-GNN) based CD} \\  \hline
    $\mathbf{U}, \mathbf{u}_s = \mathbf{u}_s^{(0)}$ & student embeddings, student $s$'s local embedding\\
    $\mathbf{V}, \mathbf{v}_e = \mathbf{v}_e^{(0)}$ & exercise embeddings, exercise $e$'s local embedding\\
    $\mathbf{O}, \mathbf{o}_k $ & concept embeddings, concept $k$'s embedding\\
    $Z$ & the dimension of latent embeddings  \\
    ${\mathcal{G}}$ & the original student-exercise graph\\
    $\tilde{\mathcal{G}}$ &  the reliable student-exercise graph  \\
    $\tilde{\mathcal{G}}_1,\tilde{\mathcal{G}}_0$& student-exercise subgraphs with different edge types   \\
    $\mu_{e\rightarrow s,\tilde{\mathcal{G}}_1}^{(l)}, \mu_{e\rightarrow s,\tilde{\mathcal{G}}_0}^{(l)}$&  edge-specific messages from $e$ to $s$ at $l$-th layer\\
    $\mu_{s\rightarrow e,\tilde{\mathcal{G}}_1}^{(l)}, \mu_{s\rightarrow e,\tilde{\mathcal{G}}_0}^{(l)}$&  edge-specific messages from $s$ to $e$ at $l$-th layer\\
    $\mathcal{N}_{s,\tilde{\mathcal{G}}_1}, \mathcal{N}_{s,\tilde{\mathcal{G}}_0}$& student $s$'s neighbor sets in different subgraphs\\
    $\mathcal{N}_{e,\tilde{\mathcal{G}}_1}, \mathcal{N}_{e,\tilde{\mathcal{G}}_0}$& exercise $e$'s neighbor sets in different subgraphs \\
    $\mathbf{u}_s^{\tilde{\mathcal{G}}}, \mathbf{v}_e^{\tilde{\mathcal{G}}}$ &  final embeddings of student $s$ and exercise $e$\\
    $<,>; \sigma$ &  inner dot and sigmoid activation\\  
    $\hat{r}_{se}^{\tilde{\mathcal{G}}}$ & predicted response log based on graph {\small${\tilde{\mathcal{G}}}$}\\
    $\mathbf{p}_{se}^{\tilde{\mathcal{G}}}$ & hidden representation based on graph {\small${\tilde{\mathcal{G}}}$}\\
    $\theta$ & all learnable parameters of S-GNN based CD\\  
    \hline
    \multicolumn{2}{c}{Informative Edge Differentiation (IE-Diff) layer} \\   \hline
    $I(X;Y)$ & mutual information between two variables \\
    ${{g}}_{se}, \tilde{{g}}_{se}$ & edge in the original/reliable graph \\ 
    $(1-{\rho}_{se})$ & edge uncertainty (probability of edge dropping) \\
    $\mathbf{W}_0$, $\mathbf{W}_1$ &  trainable weight matrices \\
    $w_{se}^0, w_{se}^1$& parameters of Bernoulli distributions\\
    $Bern(w_{se}^0), Bern(w_{se}^1)$ & Bernoulli distributions for different edges \\
    $\delta, t, \epsilon$ & random variable, temperature, noise term\\
    $X', Y'$ & two independent copies of variables $X, Y$\\
    $\mathcal{K}_X $,$\mathcal{K}_Y $ & kernel functions of variable $X,Y$ \\
    $J $ & the centering matrix \\
    $Tr(\cdot)$ & the trace of a matrix\\ 
    $\mathbf{I}$ & the identity matrix\\ 
    $\beta$ & balancing parameter between two tasks\\ 
    $\alpha$ & sharpness of kernel functions\\
    ${\mathbf{U}^{\tilde{\mathcal{G}}}},{\mathbf{V}^{\tilde{\mathcal{G}}}}$ & final embeddings based on the reliable graph\\ 
    ${\mathbf{U}^{{\mathcal{G}}}},{\mathbf{V}^{{\mathcal{G}}}}$ & final embeddings based on the original graph\\ 
    $\phi$ & all learnable parameters of IE-Diff\\ 
    \hline
    \end{tabular}}
\end{table}

\section{More Experimental Results}
\label{sec:add_exp}
\subsection{Comparisons between IE-Diff and Ada-Diff}
In this part, we focus on comparing the performance of Ada-Diff and IE-Diff. We present the results in Table \ref{transform}. 

\begin{table}[ht]
    \centering
    \caption{Comparisons between IE-Diff and Ada-Diff on the {ASSIST} and Junyi datasets. } 
    \label{transform}
    \scalebox{0.85}{
    \begin{tabular}{cc|ccc}
    \hline
        \textbf{Data} &\textbf{Model} & \textbf{ACC $\uparrow$}  & \textbf{AUC $\uparrow$} & \textbf{DOA $\uparrow$}  \\ \hline
        \textbf{ASSIST} &\textbf{Ada-Diff} & {0.7283  $\pm$ 0.0222} & {0.7555  $\pm$ 0.0244} & {0.6383  $\pm$ 0.0207} \\ 
        \textbf{ASSIST} &\textbf{IE-Diff} & {0.7322  $\pm$ 0.0247} & {0.7604  $\pm$ 0.0296} & {0.6582  $\pm$ 0.0251}  \\ \hline
        \textbf{Junyi} &\textbf{Ada-Diff} & {0.7647  $\pm$ 0.0047} & {0.7998  $\pm$ 0.0061} & {0.6484  $\pm$ 0.0146} \\ 
        \textbf{Junyi} &\textbf{IE-Diff} & {0.7672  $\pm$ 0.0041} & {0.8058$\pm$ 0.0039} & {0.6728  $\pm$ 0.0208} \\ \hline
    \end{tabular}}
\end{table}

We have two observations from Table \ref{transform}. First, Ada-Diff has a slightly worse performance than IE-Diff, though Ada-Diff has relatively superior performance than some strong baselines. 
Second, Ada-Diff necessitates an additional inference step on all response logs at each epoch. 
IE-Diff only calculates the certainty of each edge based on student and exercise local latent embeddings. This process does not require an additional GNN process. 
Third, the performance of Ada-Diff heavily relies on the manual choice of threshold {\small$\Delta$}. 
In summary, IE-Diff not only delivers superior performance metrics but also operates with a more efficient computational complexity.

\end{document}